\documentclass[aps,prd,onecolumn,10pt]{revtex4-2}

\usepackage{graphicx}
\usepackage{amsmath}
\usepackage{natbib}
\usepackage{xcolor}
\usepackage[percent]{overpic}
\usepackage{tikz}
\usepackage{url}
\usepackage[colorlinks,bookmarksopen,bookmarksnumbered,citecolor=magenta,urlcolor=red]{hyperref}
\usepackage{times}
\usepackage{epstopdf}
\usepackage[title]{appendix}
\begin{document}
\bibliographystyle{abbrvnat}
	
\title{Near-inertial waves enhance vertical transport at ocean fronts}
	
	
	
	
\author{Nihar Paul$^{a}$ and Amala Mahadevan$^{a}$}

\affiliation{$^a$Woods Hole Oceanographic Institution, Massachusetts, US}
	
	
%

\begin{abstract}
\noindent The interactions between near-inertial waves (NIWs) and submesoscale currents in the surface ocean are challenging to deconvolve due to their overlapping temporal and spatial scales. The frequency of NIW is modulated by the relative vorticity, $\zeta$, of submesoscale currents, which varies between positive and negative $\zeta$ of $O(f)$ on spatial scales of 1 -- 10~$km$, particularly across fronts where the horizontal buoyancy gradient, $\nabla_H b$, is intensified. The effective NIW frequency $f_{\scriptstyle{eff}} = f + \zeta/2$ can therefore also vary by $O(f)$ on these scales, causing the waves to be out of phase. This generates periodic convergence and divergence in the surface layer, particularly at fronts. The resulting vertical motion, known as inertial pumping, is traditionally considered to be reversible. However, the strong vertical shear of the horizontal velocity at fronts, $v_z \sim |\nabla_H b|/f$, implies that not all of the water that is pumped downward will return. We examine the effect of this asymmetry on the vertical transport of tracers with an ambient vertical gradient, analogous to biogeochemical tracers, such as oxygen and dissolved organic carbon. Using numerical simulations of an unstable front forced by NIW, we demonstrate that inertial pumping can lead to net vertical tracer transport. Spectral analysis of the vertical tracer flux given by the covariance between tracer and vertical velocity anomalies reveals that the interaction of strong NIW with submesoscale currents enhances the vertical exchange at the front on both the sub-inertial and inertial time scales.
\end{abstract}

\maketitle

\section*{Plain Language Summary}

Near-inertial waves (NIWs) are ubiquitous, can propagate long distances across the ocean basin, and affect turbulent mixing and dissipation within the mixed and topographic boundary layers. Past studies have shown the effect of NIWs on energy transfer across scales between balanced and unbalanced flow across a front. None of them highlights tracer transport in this scenario, which can influence nutrient cycling, primary productivity, and carbon sequestration in the upper ocean. In this work, we highlight and characterize subduction across a front driven by NIWs using process study modeling and observations collected during a ``Coherent Lagrangian Pathways from the Surface Ocean to Interior'' (CALYPSO) survey in the Balearic Sea during February-March 2022. Given the increase in storm events in ocean basins due to climate change, this study examines the potential role of NIWs in enhancing vertical transport by interacting with the submesoscale flow at scales of 1-10~$km$. This mechanism of enhanced transport is underestimated in coarse-resolution ocean models that are forced by smoothed wind fields.

\section{Introduction}
\label{sec:1}

The vertical transport of momentum, heat, salt, and biogeochemical tracers plays a central role in the exchange of energy between the atmosphere and ocean, in supporting the primary production of phytoplankton, and in ventilating the stratified interior ocean 
\citep{mahadevan2000modeling,mahadevan2002biogeochemical,nagai2015dominant,alford2016near,mahadevan2016impact,zhang2022vertical,paul2021eddy,paul2023eddy}. Submesoscale fronts, characterized by strong horizontal density gradients and relative vorticity, exert a strong influence on the vertical transport of nutrients, phytoplankton, carbon, and oxygen \citep{freilich2021coherent}. Frontogenesis \citep{hoskins1982mathematical} leads to the intensification of lateral buoyancy gradients and vorticity, which can give rise to vertical velocities of $\mathcal{O}$(100 $m$ $day^{-1}$) and upwelling and subduction of surface waters at oceanic fronts \citep{pollard1992vorticity,wang1993model,samelson1995evolution,spall1995frontogenesis}. More recently, field observations and simulations suggest that turbulence in the boundary layer, coupled with frontal circulation, can further enhance subduction from the mixed layer \citep{pham2024rapid}.

In addition to currents, inertia-gravity waves (IGWs) contribute significantly to the kinetic energy of the ocean. Using high-resolution numerical simulations, \cite{torres2018partitioning,torres2019diagnosing,torres2022separating} show that in certain regions and seasons, the energy levels associated with waves are comparable to or exceed those of geostrophically balanced motions at horizontal scales of 10 to 100~$km$ \citep{yang2023oceanic,thomas2023turbulent}.  
IGWs, which have frequencies between the Coriolis frequency, $f$, and buoyancy frequency, $N$, are dominated energetically by near-inertial waves (NIWs) with frequencies close to $f$, as seen in the KE spectrum. NIWs are often excited by winds, typically during storms or hurricanes, when winds impart momentum to the upper ocean \citep{pollard1970generation,kundu1986two,d1995upper_a,d1995upper_b,d1995upper_c}. They play a key role in transferring energy from the ocean surface to deeper layers, thereby contributing to the global energy budget and influencing large-scale circulation and mixing \citep{wunsch2004vertical,ferrari2009ocean}. 

IGWs can also cause vertical motion
of $\mathcal{O}$(100 $m$ $day^{-1}$), but in the linear regime, the vertical excursions experienced by water parcels are reversible and do not lead to vertical transport when integrated over time. However, net transport can occur for properties that are modified over time scales comparable to the wave period. For instance, a fraction of the nutrients uplifted into the euphotic zone by IGWs may be consumed by phytoplankton on timescales $O(1$~$day$), before the nutrient-rich isopycnals subside back to depth. Turbulent mixing also leads to irreversible transport through a variety of mechanisms, including Stokes drift and wave breaking
\citep{polton2005role} and shear-instabilities, which are modulated by ocean stratification \citep{muller1986nonlinear,polzin2008mesoscale,polzin2010mesoscale}. Nonlinear processes become especially important when wave amplitudes are large, leading to enhanced mixing through wave–wave interactions and breaking \citep{polzin2014finescale}. 

Since submesoscale currents are characterized by their large vorticity, $\zeta=O(f)$, there is a significant overlap between NIWs and submesoscale currents. A separation of NIWs and submesoscale currents can be done based on balanced {\em vs.} unbalanced motion \citep{torres2018partitioning}, as well as filtering based on frequency in the Lagrangian frame \citep{jones2023using}, but spontaneous generation of NIWs \citep{nagai2015spontaneous} and the nonlinear feedbacks between waves and currents make the separation challenging \citep{vanneste2013balance}. 

The aim of this study is to examine whether the interaction between wind-generated NIWs and submesoscale frontal flows leads to a net vertical exchange of water that exceeds the vertical transport in the absence of the waves. This is assessed through the vertical transport of a passive tracer. When viewed in isolation, waves make little or no contribution to the net advective transport. The vertical shear associated with NIWs can lead to mixing, characterized by an enhanced diffusivity at the base of the mixed layer. But the interaction between waves and currents can enhance vertical motion and induce a rectification effect, resulting in net transport not due to mixing. 

The frequency of NIWs ($\approx f$) is modulated by the relative vorticity, $\zeta$, of the background flow, to result in an effective frequency $f_{\scriptstyle{eff}}=f+\zeta/2$ \citep{kunze1985near,young1997propagation,whitt2013near}. At submesoscales, the effective frequency of NIW can differ by $O(f$) on either side of a front, since the denser side typically has positive (cyclonic) $\zeta$ and the less dense side of the front has negative (anticyclonic) $\zeta$, both of which can be $O(f)$, although typically, the cyclonic vorticity is larger than the anticyclonic vorticity at submesoscale fronts. Consequently, the NIWs have different frequencies on either side of the front, resulting in a time-varying phase difference between the wave motion in cyclonic and anticyclonic regions. This leads to periodic horizontal convergence and divergence at fronts when the waves are out of phase, resulting in periodic vertical motion, sometimes referred to as inertial pumping. 

Inertial pumping occurs when NIWs interact with a flow with spatially varying vorticity, transferring energy downward from the NIW field in the mixed layer to IGWs in the stratified pycnocline. Its effect on the vertical transport of tracers has not received much attention in the literature, and previous studies based on high-resolution model simulations with submesoscale IGWs do not show evidence for such transport \citep{balwada2018submesoscale}. However, strong lateral buoyancy gradients, $|\nabla_H b|$, where $b$ is the buoyancy, associated with submesoscale flows are in approximate thermal wind balance and generate strong vertical shear in the horizontal velocity $|{\bf u}_z| \approx f^{-1}|\nabla_H b|$. This would imply that a tracer that is moved vertically downward from the surface due to inertial pumping would experience a different horizontal transport than at the surface, and the vertical transport may not necessarily be reversible in the opposite phase of the wave divergence. In a spatially variable submesoscale flow that evolves on inertial time scales, such a rectification effect can lead to net vertical tracer transport across a strong vertical gradient. 

Recent field observations of a submesoscale frontal jet in the Alboran Sea in the western Mediterranean under highly variable wind conditions found that drifters exhibit time-dependent divergence associated with NIWs \citep{esposito2023inertial}. This study extends the simple model used in \cite{esposito2023inertial} to a fully three-dimensional submesoscale-resolving frontal flow with strong NIWs to test the hypothesis that NIWs can enhance the vertical transport at fronts. Using observations from the western Mediterranean Sea (SI Fig.~S1), we initialize the model with a semi-idealized front (SI Fig.~S2), which is forced, initially, by inertial winds (SI Fig.~S3) to generate NIWs that are allowed to evolve freely within the baroclinically unstable flow field. 

The model is used to examine the vertical transport of a passive tracer, initialized to decrease linearly with depth. The vertical shear generated by NIWs can induce diapycnal vertical transport through mixing. But, in order to isolate the advective transport mechanisms associated with the wave-front interaction, the model's vertical eddy diffusivity, $\kappa_v$, is set to a uniform and constant value of $\kappa_v = 10^{-5}$~$m^2$~$s^{-1}$ throughout the model domain. To minimize the influence of nonlinear Ekman pumping \citep{stern1965interaction,niiler1969ekman,thomas2005intensification,chen2021interaction}, the tracer field is initialized only after the wind forcing is turned off. Our analyses are for the period when the front and NIWs are unforced and decaying. 

The manuscript is organized as follows: we describe the model configuration and numerical experiments in Section~\ref{sec:2}. Section~\ref{sec:3} presents model results, where the near-surface flow and associated subduction are characterized through estimates of tracer deficit in Sections \ref{sec:3}\ref{sec:3.1},\ref{sec:3.2}, followed by probability density functions (PDFs) of vorticity, divergence, and isopycnal slope in Section \ref{sec:3}\ref{sec:3.3}. Next, we examine the isotropic wavenumber power spectral density (PSD) of horizontal and vertical kinetic energy, along with the ratio of vorticity PSD to divergence PSD (Section \ref{sec:3}\ref{sec:3.4}). The frequency–wavenumber dispersion characteristics of vertical velocity are analyzed in Section \ref{sec:3}\ref{sec:3.5}. In Section \ref{sec:3}\ref{sec:3.6}, we describe the coupled oscillations of vorticity and divergence, and Section \ref{sec:3}\ref{sec:3.7} focuses on vertical tracer covariance spectra. Finally, Section \ref{sec:4} summarizes the results and provides a broader discussion.

\section{Model configuration and initialization}
\label{sec:2}
We use the non-hydrostatic Process Study Ocean Model (PSOM) \citep{mahadevan1996nonhydrostatic_a,mahadevan1996nonhydrostatic_b} to simulate a front in a zonally periodic channel. The north-south vertical section of the model is initialized with the temperature-salinity structure of a salinity-controlled mesoscale front (measured to 250~$m$ depth during March 2022) in the Balearic Sea in the western Mediterranean Sea, shown in Fig.~S1a \citep{Calypso22dataset}. An Argo profile is used to initialize the model from 250~$m$ to 1000~$m$. The corresponding potential temperature ($\theta$, $^\circ C$), salinity ($S$, $psu$), and squared buoyancy frequency ($N^2$, $s^{-2}$) with the contours of the potential density are shown in Fig.~S1b-d. A PDF of the horizontal buoyancy gradient $M^2=|\nabla_H b|$ --- where $b=-g(\frac{\rho-\rho_0}{\rho_0})$ is the buoyancy, $\rho$ is the potential density, $g=9.81$ $m$~$s^{-2}$ is the acceleration due to gravity, and $\rho_0=1027$~$kg$~$m^{-3}$ is a reference density --- is shown in Fig.~S1e. The front is initialized with a mixed layer depth (MLD) of 25 $m$ (35 $m$) on the lighter side (heavier), and stratification of $\mathcal{O}(10^{-5})$ $s^{-2}$ typical of winter conditions of 30-60~$m$ in the Balearic Sea \citep{vargas2022seasonal} shown in Fig.~S2a,b-e. The initial condition features a small-amplitude meander with one wavelength in the zonal direction, which triggers baroclinic instability. 

The model domain is centered at 40.5$^\circ$N, where the inertial period is 18.4~$hours$ (0.768~$day$). The model domain extends 96~$km$ in the (periodic) x direction, 192~$km$ in the $y$-direction with closed walls and free-slip boundary conditions, and 1,000~$m$ in depth. The horizontal resolution is 1000~$m$, with a stretched grid in $y$ that achieves a spacing of 2~$km$ within 40~$km$ of the solid boundaries in the north and south. There are 64 vertical levels on a stretched grid with spacing ranging from 0.5~$m$ near the surface to 54~$m$ at the deeper depth. The model time step is 432~$s$. 

Three types of numerical experiments are conducted (Table~\ref{table:numexpts}). In the first case, F, the density front is allowed to evolve freely without any surface forcing. In the second and third cases, the surface is forced by an inertial wind with zonal wind stress $\tau_x=a_0sin(ft)$ and meridional wind stress $\tau_y=a_0cos(ft)$ from $day$ 5--8 for four inertial periods. The amplitude of forcing ($a_0$) is weaker (maximum of 0.03~$Pa$) in the case F + WW, and stronger (maximum of 0.05~$Pa$) in the case termed F + SW. The wind stress in the forcing time interval ($day$~5--8) is multiplied by a Tukey window with a cosine fraction of 0.1 to smoothly transition the forcing between 0 and the maximum value. Furthermore, the meridional (y-direction) profile of wind stress is tapered using a $\tanh$ profile within 48~$km$ of the northern and southern boundaries to prevent any concentrated up-/down-welling in the grid cell closest to the wall boundaries. The experiments are run for a total of 60~$days$, and the corresponding forcing profile is shown in SI Fig.~S3. 

\begin{table}[htbp]
	\centering
	\begin{tabular}{ |c | c | c | c| } 
		\hline
		Model run & F & F + WW & F + SW  \\
		\hline
		Max wind forcing & None & 0.03 Pa & 0.05 Pa \\ 
		KE$_{w}$/KE$_m$ & - & 40 & 120  \\ 
		\hline
	\end{tabular}
	\caption{Three sets of numerical experiments are conducted. Case F is just the front without any forcing. In the other two experiments, the front is initially forced by winds. Case F + WW  has weaker waves, and case F + SW has stronger waves. The wind is applied as an inertial stress over 4 inertial periods and is ramped up and down with a Tukey window. ${KE}_{w}/{KE}_m$ is the ratio of the wave to mean components of the kinetic energy right after the wind forcing is applied.}
	\label{table:numexpts}
\end{table}

The horizontal diffusion of momentum and tracers is 1 $m^2$ $s^{-1}$. After the winds are turned off (from $day$ 8 onward), the vertical diffusivity, $\kappa_v =10^{-5}~m^2s^{-1}$ throughout the domain.  During the time of wind forcing ($day$ 5--8)  $\kappa_v(z)$ is a prescribed as a function of windstress \citep{mahadevan2010rapid,he2021source} according to
$\kappa_v = max[\kappa_{max}[1 + tanh(\frac{z+\delta_e}{\Delta}], \kappa_{min} ]$, where $\delta_e = \frac{0.4}{f}\sqrt{\frac{\tau}{\rho_0}}$, is the surface Ekman depth and $\Delta = 0.5\delta_e$.  The windstress magnitude is $\tau$,  $\kappa_{min} = 10^{-3}$ $m^2$ $s^{-1}$ and $\kappa_{max} = 10^{-5}$ $m^2$ $s^{-1}$. 
The model has a flat bottom and a linear bottom drag of 10$^{-5}$ $m$ $s^{-1}$. The density is adjusted to a stable state by convective adjustment.

A tracer with concentration $c$ ($mg$ $m^{-3}$) is initialized as a linear profile ranging from 1 at the surface to 0 at the bottom (SI Fig.~S2h) after the winds are turned off on $day$~8. The model analysis is performed from $days$~8 to 60 of the model simulation.

The density front is initialized with a high local Rossby number, $Ro = \zeta/f$, defined as the ratio of vertical relative vorticity, $\zeta=v_x-u_y$, to $f$, is initially $\mathcal{O}(2)$. The frontal lateral buoyancy gradient $M^2 = |\nabla_H b|$ normalized by $f^2$  has an initial value of $M^2/f^2 =101.13$, where $f =9.45\times10^{-5}$ $s^{-2}$. This results in an initial surface rms velocity $u_{rms} = 10~ cm\: s^{-1}$ (SI Section 1). The initial Richardson number $= N^2/ U_z^2 = 0.3$ at the base of the MLD $H=25\:m$ and $N^2=2\times10^{-4}\:s^{-2}$ on the lighter side of the front.  

The maximum ratio of wave kinetic energy to mean kinetic energy $\Gamma=KE_w(\omega>0.8f)/KE_m(\omega<0.8f)$ is 40 and 120 for $|\tau|=0.03$ and $0.05~Pa$, respectively, for the front with weak waves (F+WW) and the front with strong waves (F+SW). In the following section, we compare the dynamics of the front-only case, F, with the F+WW and F+SW cases. 

One way to examine the flow is in terms of the rotational and irrotational (divergent) parts of the velocity field. The Helmholtz decomposition is used to partition the horizontal velocity $(u,v)$ into rotational $u_r,v_r$ and divergent $u_d,v_d$ components as

\begin{equation}\label{eqn:1}
	u=u_r+u_d=-\psi_y+\phi_x, \; \; v=v_r+v_d=\psi_x+\phi_y,
\end{equation}

where the stream function $\psi$ and the velocity potential $\phi$ are such that

\begin{equation}\label{eqn:2}
	\nabla_h^2 \psi = \zeta = v_x-u_y \mbox{~~and~~} \nabla_h^2 \phi = \delta = u_x+v_y.
\end{equation}

In what follows, we apply this decomposition to the kinetic energy.

\begin{figure}
	\centerline{\includegraphics[width=0.9\textwidth]{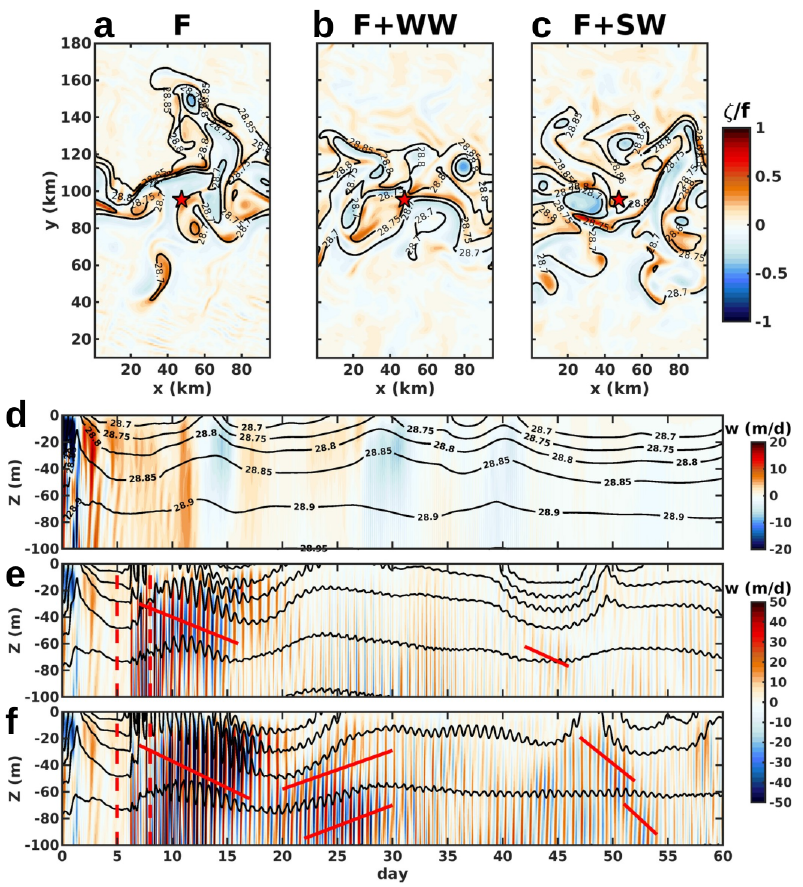}}
	\caption{Panels (a–c) show the relative vorticity, $\zeta$, normalized by $f$, overlaid with potential density contours at a depth of 3.5 $m$ on $day$~60 for the three cases: the front (F), front with weak waves (F+WW), and front with strong waves (F+SW). By $day$~60, the wave energy has greatly attenuated. Panels (d–f) are time series of the vertical velocity (\textit{w}) in the upper 100~$m$ with potential density contours (black) at the domain center marked by a star, for the same cases: (d) F, (e) F+WW, and (f) F+SW. In panels (e) and (f), the vertical dashed lines indicate $days$~5–8, during which time the inertial wind stress is applied. The solid red lines in panels (e) and (f) correspond to a group velocity on the order of 3-5 $m$~$day^{-1}$.}\label{f1}
\end{figure}

\section{Results}
\label{sec:3}
\subsection{Evolution of front and waves}
\label{sec:3.1}

Over time, the front becomes unstable, exhibiting meanders and generating eddies of 20–30~$km$ diameter, as well as filamentary structures (Fig.~\ref{f1}a-c). The vertical component of the relative vorticity, 
$\zeta$, reaches values of $O(f)$ at the front, consistent with submesoscale dynamics. In the F+WW and F+SW cases, eddies develop faster and become larger than in the case F (Fig.~\ref {f1}(a–c)). By the end of the model simulation (at 60~$days$ shown in Fig.~\ref{f1}a-c), all three experiments exhibit comparable magnitudes of vorticity.

In the F+WW and F+SW experiments, inertial wind forcing applied between $days$~5–8 excites near-inertial waves (NIW) that generate vertical oscillations (Fig.~\ref{f1}d-f). The time-series (H\"{o}vm\"{o}ller diagrams) of vertical velocity, $w$, {\em vs.} depth in the upper 100~$m$ at the domain center (Fig.~\ref{f1}d–e) show substantially higher $w$ of alternating sign in the F+WW and F+SW cases, reaching $\sim$~60~$m~day^{-1}$, compared to only 20~$m$ $day^{-1}$ in the F case. The waves lift and lower the isopycnal surfaces (black contours), in contrast to the F case, where the vertical motion is along sloping isopycnals or due to isopycnal movement \citep{freilich2019decomposition}. In the F+WW and F+SW cases, wave energy propagates downward, with a group velocity component of roughly 3–5 $m~day^{-1}$. 
Additionally, waves reflect off the bottom boundary and meridional walls, propagating upward, as shown in Fig.~\ref{f1}f. Overall, the simulations demonstrate the propagation pathways of NIWs generated by wind forcing, capturing both downward and upward energy fluxes. Such behavior is qualitatively consistent with subsurface moored observations of wind-induced internal waves interacting with mesoscale frontal features, for example, in the central Sea of Japan \citep{kawaguchi2020near}. An analysis of rotational and divergent kinetic energy using Helmholtz decomposition (SI Section 3, Fig.~S4) shows that in the case of the front (F only), the kinetic energy (KE) associated with the rotational component of the flow is much larger than the KE of the divergent component. In the wave-forced cases (F+WW, F+SW), both the rotational and divergent KE are larger than in the F-only case and diminish over time. Further, the H\"{o}vm\"{o}ller diagrams of horizontal and vertical kinetic energy are shown in SI Figs. S5 and S6, respectively, indicate that wave forcing enhances NIW energy in the upper 100~$m$.

\subsection{Tracer transport}
\label{sec:3.2}

\begin{figure}
	\centerline{\includegraphics[width=\textwidth]{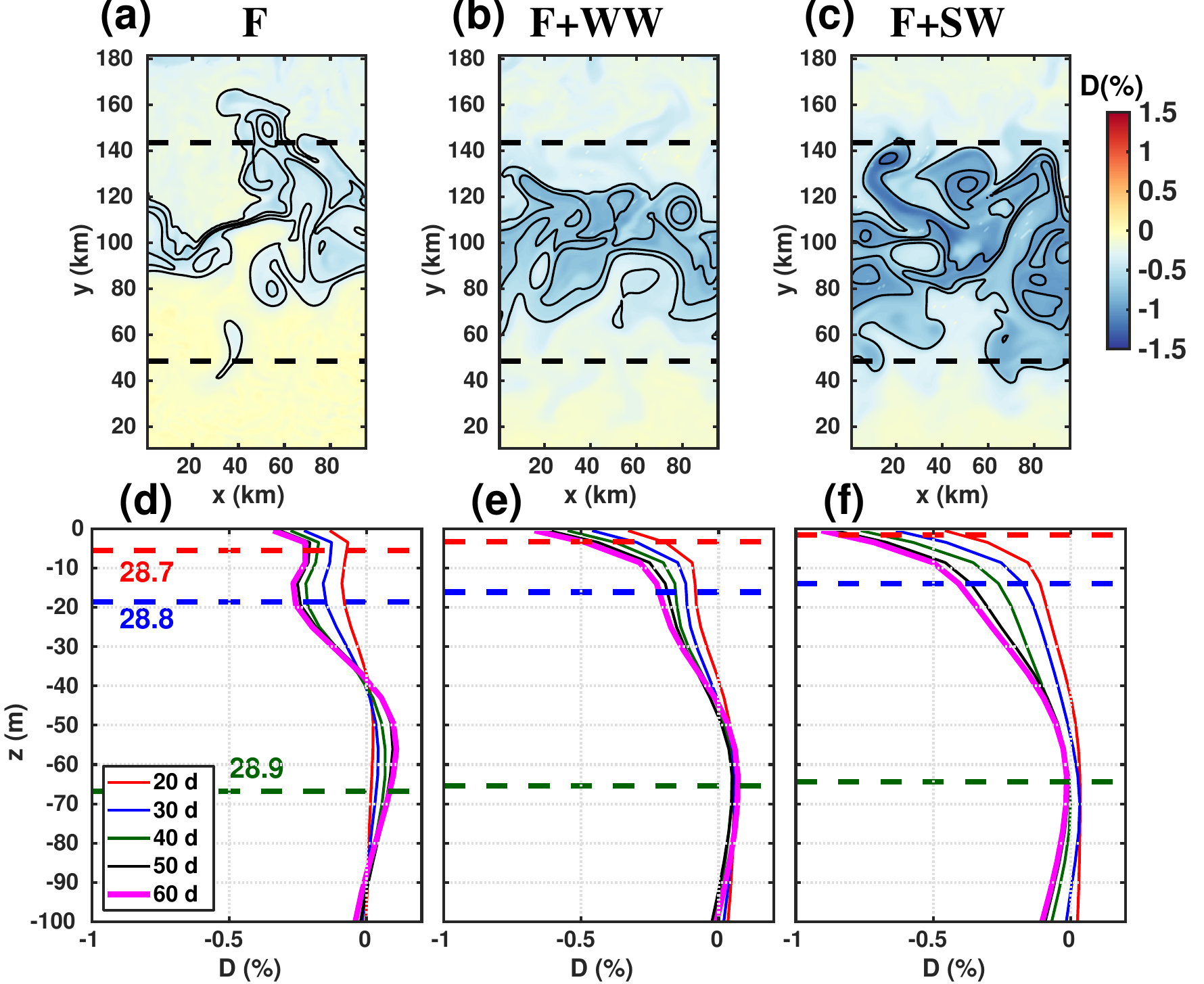}}
	\caption{Panels (a–c) show the spatial distribution of tracer deficit ($D$, in \%) on $day$~60 relative to $day$~8 when the tracer is initialized, for the three cases: Front (F), front with weak waves (F+WW), and front with strong waves (F+SW). Potential density anomaly contours at 3.5~$m$ are overlaid in each panel for $day$~60. The horizontal dashed lines indicate $y=48~km$ and 144~$km$, which mark the meridional extent of the region used for horizontal averaging. Panels (d–f) horizontally averaged vertical profiles of the tracer deficit, $D$\%, over the upper 100~$m$. The average depths of the isopycnals 28.7 (red), 28.8 (blue), and 28.9 $kg~m^{-3}$ (green) on $day$~60 are shown in panels (d-f).}\label{f2}
\end{figure}

Most biogeochemical properties exhibit vertical gradients in the upper ocean. We therefore use an idealized passive tracer, $c(x,y,z,t)$, initialized on $day$~8 when the wind stress is turned off. The initial tracer distribution, $c_i(z)$, is horizontally uniform and varies linearly in $z$, ranging from 1 at the surface to 0 at the seabed. The normalized tracer deficit, $D=(c(x,y,z,t)-c_i(z))/c_i(z)$ is indicative of cumulative vertical transport (Fig.~\ref{f2}). A comparison of the horizontal distribution of $D(x,y,z,t)$ near the surface ($z=-3.5~m$) between the three model runs F only, F+WW, and F+SW in (Fig.~\ref{f2}a--c), shows that NIWs enhance the vertical transport in the frontal zone, where horizontal buoyancy gradients are present. The vertical transport of tracer is largest for the case F+SW. Away from the frontal zone, the waves make little or no impact on vertical transport.

The tracer deficit ($D$), averaged zonally and meridionally between $y=~48$ and 144~$km$ (Fig.~\ref{f2}d–f), indicates that tracer anomalies in the upper 100~$m$ increase over time as the tracer is transported down-gradient from the surface to the interior. Because the tracer is conserved in these model runs, its depletion from the upper ocean is compensated by accumulation at depth. The upward transport of a tracer whose concentration increases with depth can be calculated as $1-c$. A comparison across model runs reveals that when NIWs interact with the front, as in the F+WW and F+SW cases, the vertical tracer transport from the upper 10–25~$m$ is enhanced by a factor of 3–4 relative to case F. By $day$ 60, the average depths of the isopycnals $\sigma_\theta = 28.7, 28.8$, and 28.9 $kg~m^{-3}$ in the wave-forced cases F+WW and F+SW is slightly less than the case F (Fig.~\ref{f2}d-f). We assess the tracer transport across the approximate depths of these isopycnals, i.e.,  $z=3.5$~$m$, 20~$m$, and 60~$m$.  
From tracer-transport analysis, we conclude that tracer vertical transport is substantially increased when both the front and NIWs are present, compared to the other cases (SI Movie S1). 

\begin{figure}
	\centerline{\includegraphics[width=0.95\textwidth]{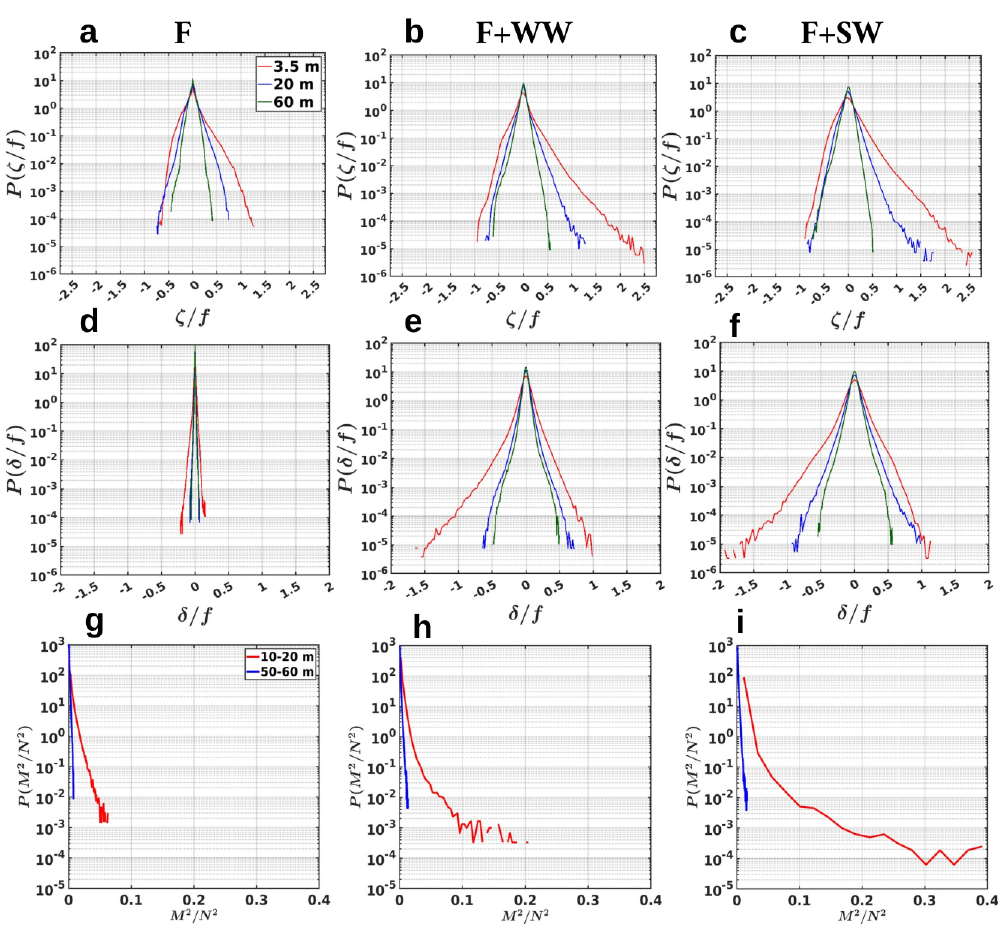}}
	\caption{Probability density functions (PDFs) of (a–c)  normalized relative vorticity $\zeta/f$; (d–f) horizontal divergence $\delta/f$, where $\delta= u_x +v_y$; and (g–i) the ratio of horizontal buoyancy gradient magnitude to vertical stratification, $M^2/N^2= |\nabla_H b|/(\partial b/\partial z~$), with $b$ denoting buoyancy.. The left column is the front-only case (F), the middle column -- front with weak waves (F+WW), and the right column -- front with strong waves (F+SW). The PDFs are computed over the period between $day$~8 and $day$~60, after the winds are turned off, at depths of 3.5~$m$ (red), 20~$m$ (blue), and 60~$m$ (green) for the upper two panels and 10-20 $m$ and 50-60 $m$ for the lower panel, respectively.}\label{f3}
\end{figure}

\subsection{Vorticity, divergence, and isopycnal slope}
\label{sec:3.3}

Submesoscale dynamics is known to locally intensify the lateral buoyancy gradient $M^2= |\nabla_H b|$ through frontogenesis \citep{mcwilliams2016submesoscale,barkan2019role}, as well as the relative vorticity $\zeta$ and divergence $\delta=u_x+v_y$ in the near-surface layer \citep{shcherbina2013statistics,balwada2021vertical}. 
The PDFs of $\zeta$, $\delta$, and isopycnal slope $M^2/N^2$ (Fig.~\ref{f3}) for the F, F+WW, and F+SW cases highlight the characteristic intensification of these quantities near the surface (3.5~$m$), at 20~$m$, and 60~$m$. The PDFs are computed from $days$~8 to 60, after the wave-forced cases' inertial wind forcing is turned off. Within the mixed layer, the distribution of relative vorticity is positively skewed, with filaments of strong cyclonic vorticity embedded in a weaker, predominantly anticyclonic background. This positive skewness decreases with depth. In contrast, the PDF of $\delta$ is negatively skewed, reflecting stronger convergence, with the skewness weakening with depth. 

In particular, the presence of waves in the F+WW and F+SW cases substantially enhances the relative vorticity and its positive skewness (Fig.~\ref{f3}b,c), as well as the divergence and its negative skewness (Fig.~\ref{f3}e,f). The increased occurrence of large positive vorticity and negative divergence in these cases is consistent with larger isopycnal slopes, reflected by higher values of 
$M^2/N^2$, relative to the F-only case. This indicates that NIWs not only locally amplify vorticity and divergence at the front but may also intensify frontogenetic processes \citep{hoskins1982mathematical}. Consequently, the enhanced vertical velocities arise from a combination of wave-induced motions and strengthened secondary circulations at the fronts. 

\begin{figure}
	\centerline{\includegraphics[width=\textwidth]{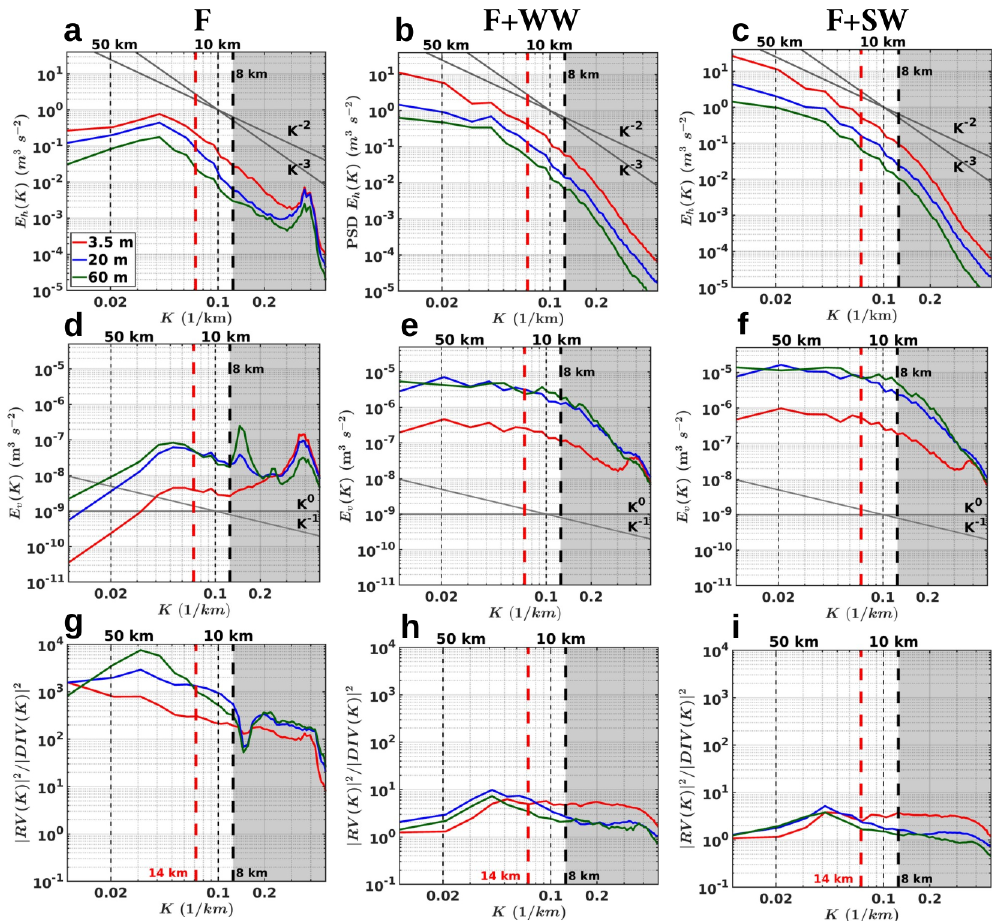}}
	\caption{Panels (a–c) show the isotropic power spectral density (PSD) of horizontal kinetic energy, (d–f) vertical kinetic energy, and (g–i) the ratio of PSD of relative vorticity (RV) to PSD of divergence (DIV), all averaged over $days$~8–60. The spectra are plotted as a function of horizontal wavenumber ($K=\sqrt{k^2_x+k^2_y}$, where $k_x$ and $k_y$ are the zonal and meridional wavenumbers, respectively, at depths of 3.5~$m$ (red), 20~$m$ (blue), and 60~$m$ (green). Results are shown for the three cases: the front (F), the front with weak-wave forcing (F+WW), and the front with strong-wave forcing (F+SW). Reference slopes of \{-3, -2\} for horizontal kinetic energy (a–c) and \{-1,0\} for vertical kinetic energy (d–f) are indicated. The gray-shaded region to the right of 8~$km$ (thick black vertical dashed line) marks the unreliable part of the PSD, where the model resolution could impose limitations. The red dashed line denotes the internal Rossby radius of deformation ($L_R$).}\label{f4}
\end{figure}

\subsection{Power spectral density (PSD) of horizontal KE,  vertical KE, and $PSD(\zeta)/PSD(\delta)$}
\label{sec:3.4}

The isotropic PSDs of horizontal kinetic energy (KE) are calculated within the domain along the zonal direction and 48~$km$ from the north-south boundaries, that is, within the central $96\times96$~$km^2$ of the model domain. The velocities are tapered in the y-direction using a Tukey window with a cosine fraction of 0.1, and the area mean is removed at each time step. Then, consecutive isotropic spectra are calculated following the methodology described in Section 4 of the SI. The results show spectra at depths of 3.5~$m$, 20~$m$, and 60~$m$, calculated at each timestep and then averaged over $day$~8-60.

The isotropic PSDs for the horizontal KE (Fig.~\ref {f4}a-c) and the vertical KE (Fig.~\ref{f4}d-f) reveal differences between the frontal flow (F) and the fronts with NIWs (F+WW and F+SW). The energy levels are considerably higher in the front with NIWs. The PSD of the KE associated with $u,v$ rolls off from $E_h(K) \sim K^{-2}$ to $E_h(K) \sim K^{-3}$ with $K$ (Fig.~\ref{f4}a-c), where, $K=\sqrt{k^2_x+k^2_y}$ is the isotropic wavenumber, and $k_x$ and $k_y$ are the zonal and meridional wavenumber, consistent with previous studies of quasi-geostrophic turbulence and submesoscale currents \citep{capet2008mesoscale}. The spectrum of the vertical KE, $E_v = \frac{1}{2}w^2$ is much flatter than the horizontal KE and steepens from $E_v(K) \sim K^{0}$ to $E_v(K) \sim K^{-1}$ with increasing $K$ (Fig.~\ref{f4}d-f). Unlike the horizontal KE, the vertical KE is higher below the surface. Even though we expect the divergence $u_x +v_y$ to be highest near the surface, the vertical velocity $w(z) = \int_{z}^{surface} -(u_x + v_y) dz$ peaks subsurface. The vertical KE is almost two orders of magnitude higher in the cases with NIWs, particularly in the F+SW case.  

All the spectra constructed from our model exhibit a roll-off and spurious fluctuations at high wave numbers. The case F shows spurious fluctuations beyond $K=1/8$~$km^{-1}$, suggesting that the model does not properly represent energy at horizontal scales below 8~$km$, as it lacks a subgrid turbulence closure scheme beyond a small, uniform eddy viscosity. The F+WW and F+SW cases show spurious fluctuations or roll-off beyond $K=1/4$~$km^{-1}$, but we chose a scale cut-off of 8~$km$ (shaded gray in Fig.~\ref{f4}) for consistent comparison between model runs. Furthermore, the deformation radius ($L_R$) is approximately $L_R=NH_m/f=14~km$, where $H_m$ is the depth of the mixed layer (the depth corresponding to the maximum $N$, where $N$ is calculated as the average of all grid points within the domain of the isospectra from $day$ 8 to $day$ 60), and this number is in good agreement with altimetry estimates in the Balearic Sea \citep{mason2013multiscale}. 

Submesoscale velocities are comprised of vortical geostrophic and ageostrophic motions, and unbalanced (divergent) IGWs \citep{lien1992normal,callies2013interpreting,callies2019some}. The divergence includes contributions of both submesoscale ageostrophic motions and IGWs. The PSD of vorticity and divergence (SI Fig.~S7) for the F, F+WW, and F+SW cases shows that waves slightly increase the power in vorticity and significantly increase the power in divergence. 

The spectrum of half of the vertical vorticity squared, also defined as the enstrophy, represents the geostrophically balanced part of the KE, and the square of the divergence yields the unbalanced dynamics. The power spectrum of relative vorticity, $RV=\zeta/f$, can be considered as the vortical contribution to the KE given by $|RV(K)|^2=\frac{K^2|\psi(K)|^2}{f^2}$, where $K^2|\psi(K)|^2$ is the vortical component of the KE spectrum. Using a similar argument, the power spectrum of the lateral divergence, $DIV=\delta/f$, yields $|DIV(K)|^2=\frac{K^2|\phi(K)|^2}{f^2}$, where $K^2|\phi(K)|^2$ is the divergent component of the KE spectrum. Here, $\psi$ and $\phi$ are the streamfunction and potential, used to decompose the flow into rotational and divergent components, as described in equations (\ref{eqn:1}) and (\ref{eqn:2}). Waves significantly decrease the ratio of their PSDs, i.e., $|RV(K)|^2/|DIV(K)|^2$ (Fig.~\ref{f4}g-i) as they make an outsize contribution to horizontal divergence. 

In the context of NIWs, energy transfer across scales can reduce the inverse KE cascade of the low-passed eddying flow and enhance its forward cascade. At larger spatial scales, the forward KE cascade is accomplished through wave scattering and direct extraction by rotational eddy motions. In comparison, on smaller spatial scales, it is also dominated by wave‐wave interactions \citep{shaham2025spectral}. In the given scenario, the nonlinear interaction between NIWs and eddies can result in distinct spectral characterizations in the wavenumber-frequency space within and below the initial mixed layer, as examined in the following section.

\begin{figure}[ht]
	\centerline{\includegraphics[width=\textwidth]{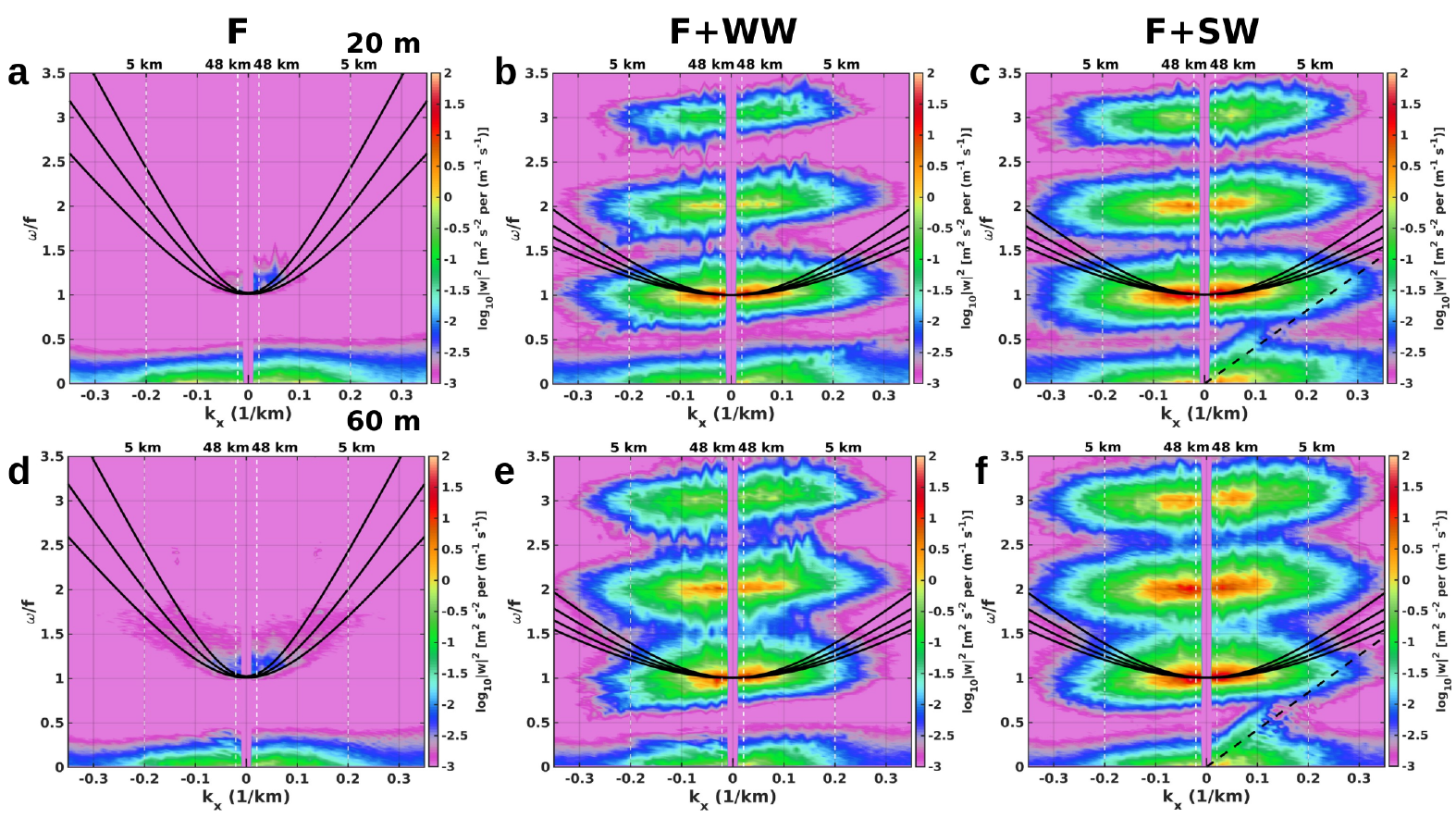}}
	\caption{Panels (a–c) and (d–f) show the zonal wavenumber ($k_x$)–frequency ($\omega/f$) power spectral density of vertical velocity, meridionally averaged, for the F, F+WW, and F+SW experiments at 20~$m$ (top row) and 60~$m$ (bottom row), over $days$~8–60. Overlaid dispersion curves correspond to channel mode $n=8$, with baroclinic phase speeds ($c_m$) for modes $m=3-5$ in F (upper to lower curves), and modes $m=7-10$ in F+WW and F+SW (upper to lower curves). In the F+SW case, the black dashed line marks the baroclinic phase speed of the Kelvin wave along the channel boundary, empirically fitted as 0.063~$m$ $s^{-1}$.}\label{f5}
\end{figure}

\subsection{Frequency-wavenumber iso-spectra of vertical velocity}
\label{sec:3.5}

The solid walls at the northern and southern boundaries of the domain inevitably support channel modes in the internal gravity wave field \citep{gill2016atmosphere}. To evaluate their influence, as well as the distribution of wave energy across dispersion characteristics, we compute the vertical velocity zonal-wavenumber ($k_x$)-frequency ($\omega/f$) power spectral density for the F, F+WW, and F+SW experiments (Fig.~\ref{f5}), evaluated at 20~$m$ (top row) and 60~$m$ (bottom row). First, the variables are tapered in time, and then the time-average of the variable at each grid point is subtracted to obtain anomalies. The power spectral density is computed using a 2D FFT along the time and zonal directions, and the result is averaged meridionally for both depths.

The dispersion relation for Poincar\'{e} internal waves in a channel of finite depth $H$ and finite width $W$, subject to no-normal-flow meridional boundary conditions \citep{gill2016atmosphere}, is given by 

\begin{equation}\label{eqn:3}
	\omega^2=f^2+c^2_mK^2=f^2+c^2_m(k^2_x+n^2\pi^2/W^2),
\end{equation}

where $\omega$ is the frequency, $W=192~km$ is the channel width, $k_x$ is the zonal wavenumber, $n$ is the meridional mode number (an integer), and $c_m$ is the baroclinic wave speed of vertical mode $m$. The term 
$n\pi/W$ represents the meridional wavenumber imposed by the channel boundaries, giving rise to discrete meridional (channel) modes, while $k_x$ remains continuous in the zonal direction. Each vertical mode~$m$ has its own baroclinic wave speed $c_m$, which governs the horizontal propagation of the waves. The dispersion curves for $m=3,4,5$ overplotted in Fig.~\ref{f5} suggest that these channel modes are present in the model simulations. 

Stratification profiles, $N^2(z)$, are computed from time-averaged densities at each grid point between $days$~8 and 60 across the domain. These profiles are then used as input to the Sturm–Liouville problem \citep{muller2015toward,kelly2015internal,kelly2016vertical} derived from the linearized Boussinesq equations with $f$ at 40.5$^\circ$N and a water depth of 1000~$m$. The corresponding, horizontally averaged $\overline{N^2(z)}$, along with the barotropic ($m=0$) and the first ten baroclinic wave speeds ($m=1-10$), are shown in SI Fig.~S8, while the associated vertical eigenfunctions are presented in Fig.~S9 for the three experiments. Stratification is weaker in the F+WW and F+SW cases compared to F alone. Consequently, the baroclinic wave speeds are slightly reduced relative to F. However, increasing the vertical grid resolution may produce greater differences in the estimated baroclinic wave speeds (see SI Table~1).

Even though experiment F is unforced, frontal adjustment leads to some spontaneous generation of NIWs, with the energy spectral density showing distinct subinertial and inertial energy packets. We focus on channel mode $n=8$, corresponding to a meridional length scale $l_y=2W/n=48$ ~$km$ \citep{gill2016atmosphere}. As shown in Fig.~\ref{f5}a,d for depths of 20~$m$ and 60~$m$, the energy packets in the F experiment are concentrated within vertical modes $m=3-5$. In contrast, in the wave-forced cases (F+WW and F+SW; Fig.~\ref{f5}b,e, and Fig.~\ref{f5}c,f), the energy packets are more dispersed along the zonal wavenumber axis and are shifted toward higher vertical modes ($m=7-10$). Additionally, both wave cases show changes in power at subinertial frequencies and the scattering of energy into superinertial harmonics ($2f$, $3f$, $\ldots$). We also empirically fit a line corresponding to the baroclinic wave speed of 0.063~$m$ $s^{-1}$, which represents the effect of the boundary-trapped Kelvin wave that decays offshore within a Rossby radius of deformation. Overall, the scattering of NIW energy to higher frequencies is consistent with idealized quasi-geostrophic (QG) simulations by \cite{dong2023geostrophic} and global-scale numerical simulations by \cite{raja2022near}, indicating nonlinear wave–wave interactions that facilitate energy transfer across scales. The following section examines how the enhanced vertical velocities induced by these waves interact with submesoscale currents to generate convergence and divergence. 

\begin{figure}
	\centerline{\includegraphics[width=\textwidth]{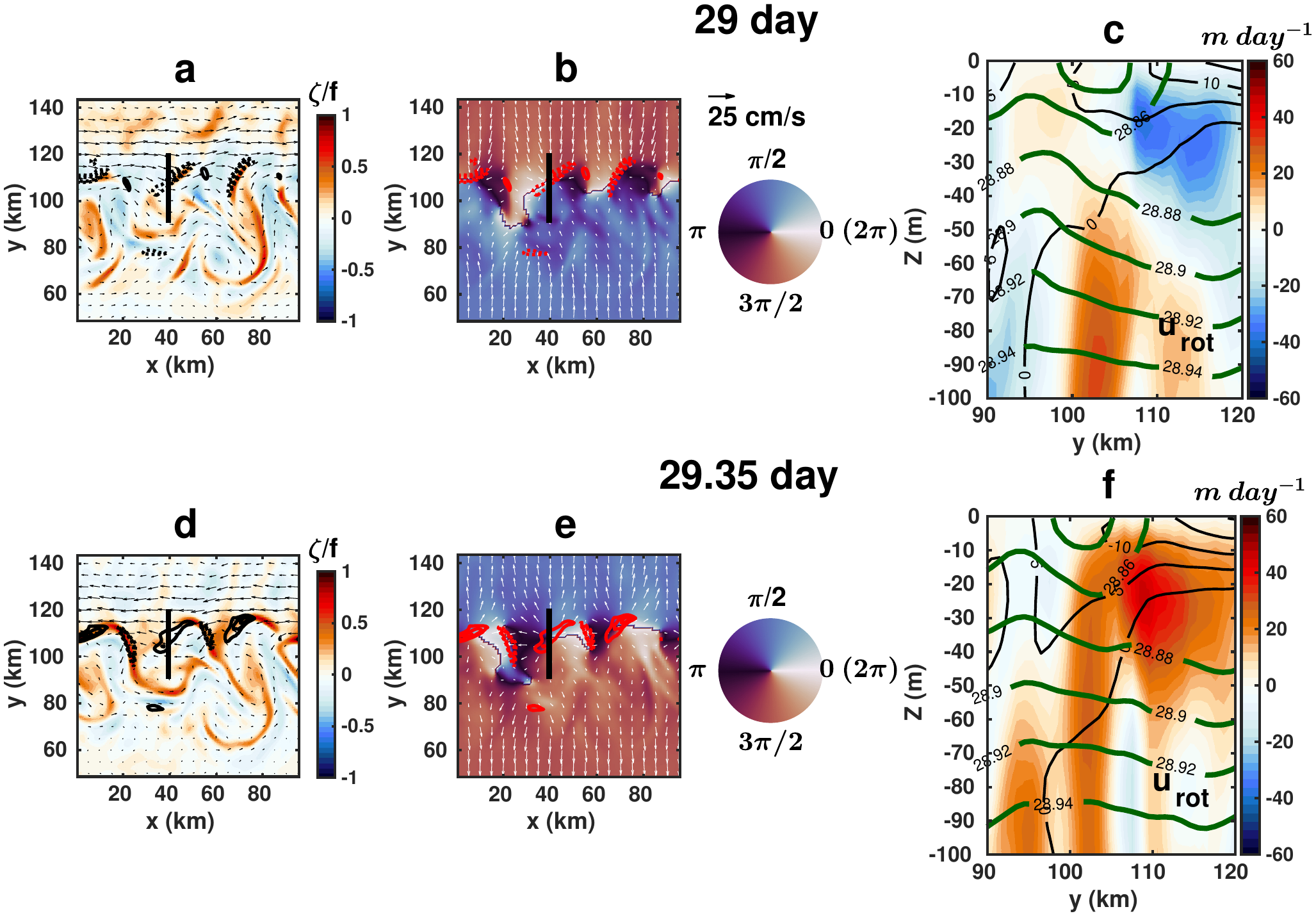}}
	\caption{Snapshots of model fields for the F+SW case at 3.5~$m$ depth on $day$~29 (top row) and $day$~29.35 (bottom row), separated by approximately half an inertial period. (a,d) Relative vorticity normalized by Coriolis frequency ($f$) overlaid with rotational velocity vectors. (b,e) Phase (color), defined as $\theta=tan^{-1}(v_d/u_d)$, where $u_d$ and $v_d$ are the divergent velocity components; contours of convergence/divergence normalized by $f$ ($|\delta/f|>0.3$) are overlaid, with convergence shown by dotted lines and divergence by solid lines. (c,f) Vertical sections along the thick black dashed line showing rotational zonal velocity ($u_{rot}$, black contours) and potential density (green contours), with vertical velocity indicated by shading over the upper 100~$m$.}
	\label{f6}
\end{figure}

\begin{figure}[ht]
	\centerline{\includegraphics[width=\textwidth]{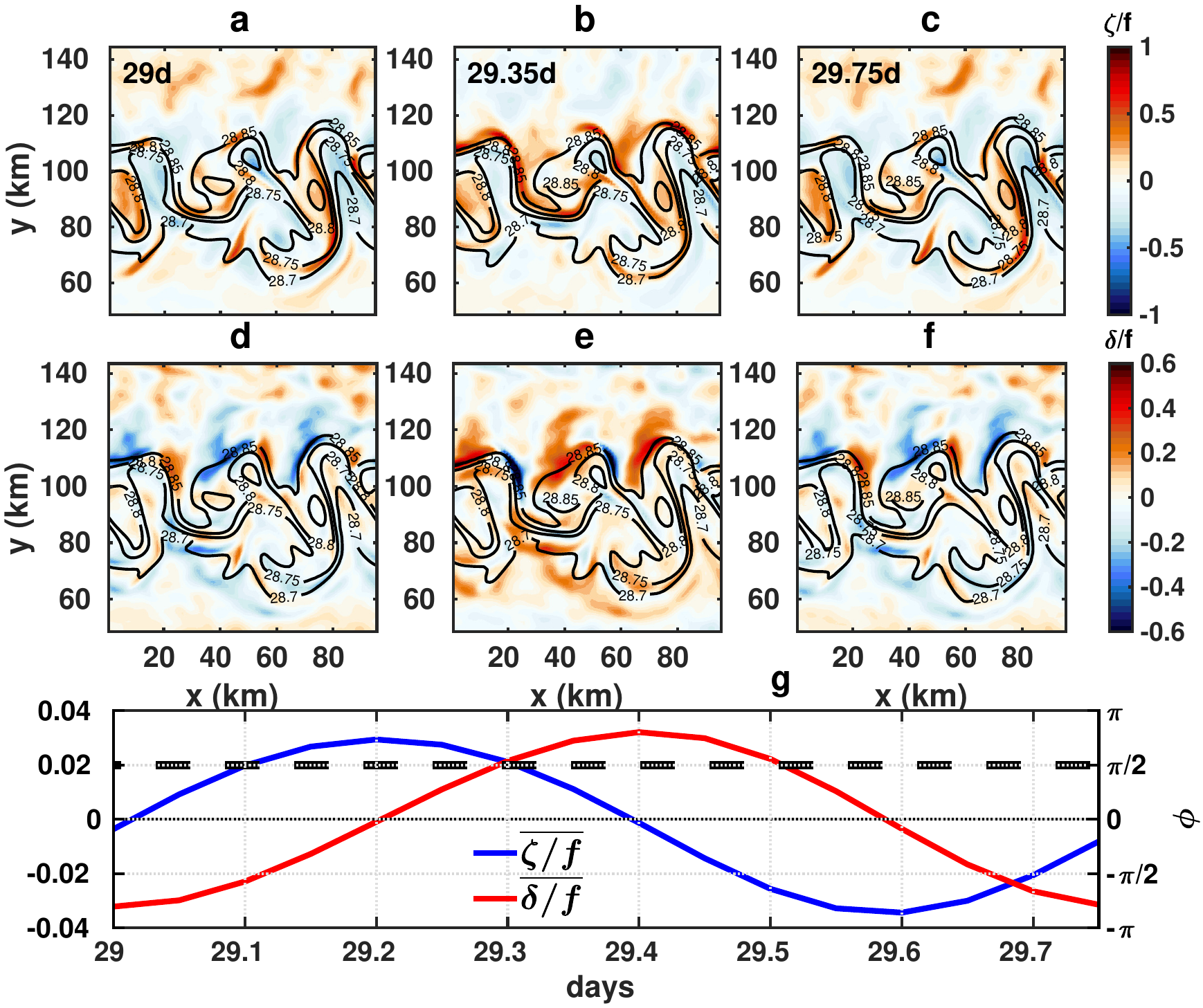}}
	\caption{Panels (a–c) and (d–f) illustrate the relative vorticity ($\zeta$) and divergence ($\delta$), respectively, normalized by the Coriolis parameter $f$, at three distinct time steps within an inertial period. These snapshots capture the frontal structure, characterized by intense wave activity at a depth of 3.5~$m$. The time intervals span from $day$~29.00 to $day$~29.75 of the simulation, with an approximate 9-hour increment between each. Overlaid contours of potential density anomaly are provided for additional reference. Panel (g) shows the spatially averaged vorticity and divergence, zonally and meridionally averaged over a 48~$km$ region from the north-south boundary, during the inertial period. Also presented is the relative phase difference $\phi$ in the range $[-\pi, \pi]$ between vorticity and divergence.}\label{f7}
\end{figure}

\subsection{Coupled oscillations of vorticity and divergence}
\label{sec:3.6}

The vortical component of the flow field is traditionally considered as comprised of the meso- and submeso-scale currents, but not the waves. However, we find that the vorticity of the flow field is strongly enhanced and modulated by NIWs (Fig.~\ref{f3}a-c), and the cyclonic relative vorticity can locally attain values significantly larger than $f$, particularly in the cases F+WW and F+SW. Concurrently, waves also increase the convergence (Fig.~\ref{f3}d-f) and vertical velocities (Fig.~\ref{f1}d-f). This can be explained as follows. 

The relative vorticity of the flow modifies the effective NIW frequency as
$f_{eff}=f+\zeta/2$. As a result, the flow's vorticity gradient results in regionally different values of $f_{eff}$. This induces a phase difference between NIWs across different regions of the flow, leading to convergence and divergence where vorticity gradients are large. Typically, fronts are associated with strong along-front currents, which give rise to a cross-front gradient in relative vorticity with cyclonic relative vorticity on the denser side and anticyclonic relative vorticity on the lighter side of the front. Hence, the effective NIW frequency differs on either side of a front, and the resulting phase difference leads to periodic, wave-driven convergence and divergence at the front (Fig.~\ref{f6}). When the two velocity vectors are in opposite phases, either directed toward or away from each other, the resulting convergence or divergence drives vertical motion within the evolving front. Convergence, where the vectors are directed inward, promotes subduction, while divergence, with the vectors directed outward, favors obduction (Fig.~\ref{f6}c,f). 

In the model simulations, we observe that initially, the NIWs exhibit phase differences that coincide with the gradients in relative vorticity. However, over time, most of the ocean comes into resonance with the inertial frequency. The smaller wave numbers dominate, and the domain is occupied predominantly by waves differentiated by approximately two opposite phases (Fig.~\ref{f6}b,e) that give rise to modulations in divergence at the inertial frequency (Fig.~\ref{f7}).

The inviscid and barotropic approximation for the vertical component of the relative vorticity, $\zeta=v_x-u_y$, obtained by taking the curl of the horizontal momentum equations, is  

\begin{equation}\label{eqn:4}
	\frac{D\zeta}{Dt} \approx \underbrace{-\delta(\zeta+f).}_{vortex\: stretching}
\end{equation}

The domain-averaged material derivative of vorticity and the right-hand side of Equation \ref{eqn:4} exhibit an approximate balance at a depth of 3.5~$m$ (SI Fig.~S10a, days 28–35 for the SW case). Moreover, regions of enhanced $D\zeta/Dt$ are associated with subduction. A coherent correspondence between $D\zeta/Dt$ and the right-hand side of Equation \ref{eqn:4} is evident in the spatial maps shown in SI Fig.~S10b–d and S10e–g over the interval from $days$~29 to 29.75. The difference between the two estimates is approximately an order of magnitude smaller than the individual terms, indicating good agreement with the proposed balance (SI Fig. S10h-j). Therefore, divergence of the horizontal velocity, $\delta = u_x +v_y$, intensifies $\zeta$ by the vortex stretching mechanism.

Conversely, the wave-driven ageostrophic relative vorticity affects the divergence $\delta$, which is obtained by taking the horizontal divergence of the horizontal momentum equations, and subtracting the geostrophic balance $fv^g = \frac{1}{\rho_0} \frac{\partial p}{\partial x}$ and $fu^g = -\frac{1}{\rho_0} \frac{\partial p}{\partial y}$, where the horizontal velocity components are 
$u= u^g + u^a$ and $v= v^g + v^a$. The evolution of $\delta$ is described by

\begin{equation}\label{eqn:5}
	\frac{D \delta}{D t} +u_x^2 + v_y^2 +2 v_x u_y  = f \zeta^a,  
\end{equation}

where $\zeta^a = v^a_x - u^a_y$ is the ageostrophic relative vorticity. Neglecting the nonlinear terms in (\ref{eqn:4}) and (\ref{eqn:5}) gives the coupled equations

\begin{align} 
	&\partial \delta/\partial t = f \zeta \label{eqn:6}\\
	&\partial \zeta /\partial t = - \delta f. \label{eqn:7}
\end{align}

The equations examine just the wave-driven part of $\zeta$, and we have dropped the superscript $a$ to denote the ageostrophic component. Also, it is assumed that for most of the ocean $\zeta \ll f$, even though locally, at submesoscale fronts $\zeta = O(f)$. Equations (\ref{eqn:6}) and (\ref{eqn:7}) represent a coupled oscillation described by

\begin{equation}\label{eqn:8}
	\frac{dZ}{dt} = if (\zeta+i \delta), \mbox{~~where~} Z = \zeta + i \delta \mbox{~~and~} i = \sqrt{-1},
\end{equation}

The solution

\begin{equation}\label{eqn:9}
	Z= e^{ift} = \cos ft + i \sin ft.
\end{equation}

shows that over the region, $\zeta = \cos ft $ and $\delta = \sin ft$, oscillate at the inertial frequency and are phase shifted by $\pi/2$ (Fig.~\ref{f7}). 
This results in the periodic modulation of $\zeta$ and $\delta$ that is 90$^\circ$ out of phase. Since $f$ dominates over $\zeta$ in many regions of the flow, the inertial frequency dominates the periodicity of  $\zeta$ and $\delta$. The alternating horizontal convergence and divergence near the surface generate downwelling and upwelling. Its effect on vertical tracer transport is assessed by examining the covariance between the vertical velocity and tracer anomalies in the following section.

\begin{figure}
	\vspace{2cm}
	\centerline{\includegraphics[width=\textwidth]{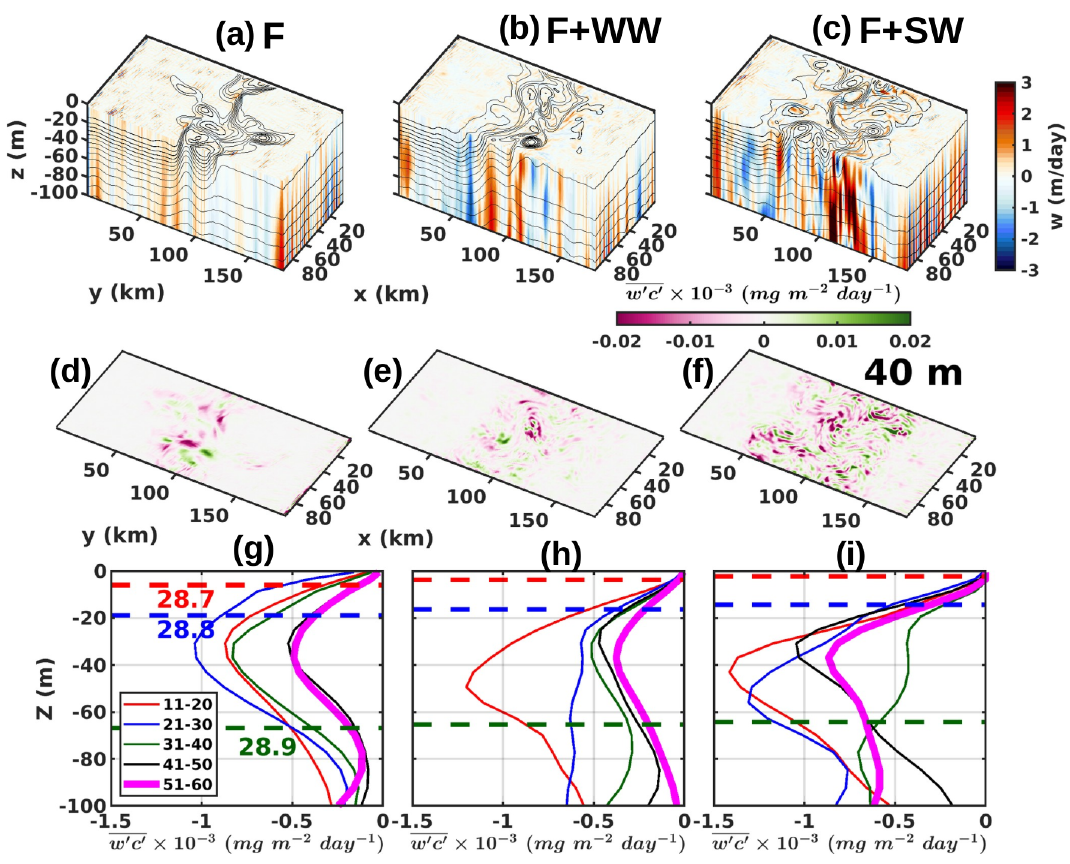}}
	\vspace{0.5cm}
	\caption{Comparison of vertical tracer fluxes for F, F+WW, and F+SW model runs are shown in each column. Panels (a–c) show the vertical velocity averaged over $days$~51–60 for the upper 100~$m$ (with the upper surface corresponding to –3.5~$m$), overlaid with contours of potential density anomaly. Panels (d–f) Covariance $\langle w'c' \rangle$ at a depth of 40 $m$ averaged over $days$~51-60. The downward tracer flux (pink) exceeds the upward flux (green). Panels (g–i) Vertical tracer flux as a function of depth, area-averaged and time-averaged over 10-$days$ intervals between $days$~11–60. The depths of the isopycnals 28.7 (red), 28.8 (blue), and 28.9~$kg~m^{-3}$ (green) averaged over $days$~51-60 relative to $z=-3.5~m$ are indicated for each model run.}\label{f8}
\end{figure}

\subsection{Vertical tracer flux}
\label{sec:3.7}

\begin{figure}
	\centerline{\includegraphics[width=0.8\textwidth]{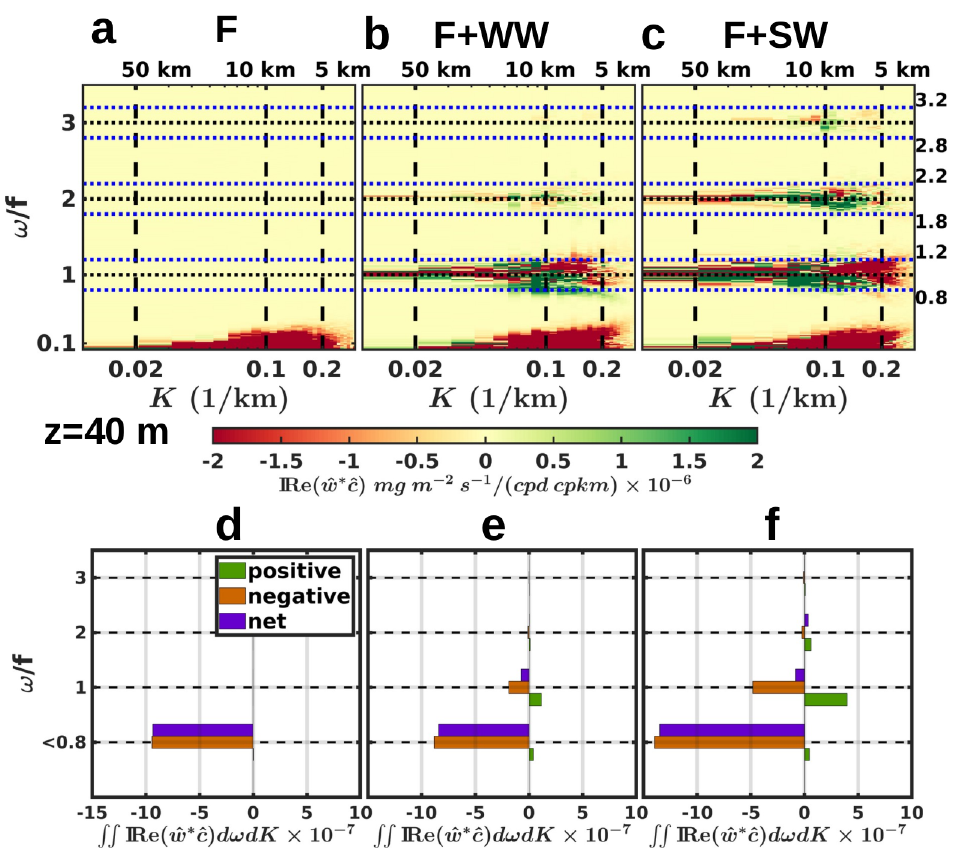}}
	\caption{(a-c) Isotropic vertical tracer-covariance spectra of tracer concentration and vertical velocity anomalies ($c^\prime$; $mg$ $m^{-3}$, $w^\prime$; $m\:s^{-1}$) in frequency-wavenumber space ($\omega-K$), where $K=(k^2+l^2)^{1/2}$, $\omega$ is the frequency, and $k,l$ are the zonal and meridional wavenumbers for the three experiments with F, F+WW, and F+SW, respectively, at a depth of 40~$m$ from 8 to 60 $days$. The horizontal dotted lines indicate the Coriolis frequency $f$ and its super-inertial frequencies ($2f$ and $3f$). The thin blue dotted lines correspond to $\pm0.2f$ about $f,2f,3f$, respectively. The vertical dashed lines correspond to horizontal scales of 50~$km$, 10~$km$, and 5~$km$. (d-f) Bar graph of integrated power for the frequency range below $0.8f$ (sub-inertial band), between $f\pm0.2f$ (inertial band), $2f\pm0.2f$, and $3f\pm0.2f$ (super-inertial bands) for the positive (green), negative (red), and net contribution (purple) from the numerical experiments.}\label{f9}
\end{figure}

The vertical tracer flux is defined as the product of vertical velocity $w^\prime$ and tracer concentration $c^\prime$ anomalies. Using the overbar to denote an area average at each timestep, and writing $c= \overline{c} + c'$ and $w= \overline{w} + w'$, with $\overline{w'}=0$ and $\overline{c^\prime}=0$, we can write 

\begin{equation}\label{eqn:10}
	\overline{w c}
	= \overline{(\bar{w} + w')(\bar{c} + c')}
	= \bar{w}\,\bar{c} + \overline{w'c'}.
\end{equation}

The covariance of $w'$ and $c'$ therefore determines the vertical tracer flux. Since the background initial tracer concentration $\overline{c}(z,t)$ decreases with depth, a downward velocity induces a positive tracer anomaly $c'$ and upward velocity is associated with a negative $c'$, so that  $w'c'<0$ contributes a net downward transport of tracer, which weakens the tracer's vertical gradient. Conversely, $w'c'>0$ is an upward tracer flux that acts to restore the tracer gradient. IGWs enhances the alternating vertical velocities averaged in time (over 13 inertial periods in Fig.~\ref{f8}a-c) for the F+WW and F+SW cases. Concurrently, the time-averaged vertical flux $\langle w'c' \rangle$ is also intensified by the presence of IGW in comparison to the case F without waves (Fig.~\ref{f8}d-f). While $\langle w'c' \rangle$  can be both positive or negative, it is predominantly negative, resulting in a net downward flux of tracer. Hence, the area- and time-averaged vertical tracer flux is negative (downward) in the upper 100~$m$ (Fig.~\ref{f8}g-i) with a maximum at approximately 40~$m$ depth. 

We define the space–time cross-covariance as

\begin{equation}\label{eqn:11}
	R_{wc}(r,\tau) = \overline{\langle w'(\mathbf{x},t)\,
		c'(\mathbf{x}+ \mathbf{r},\, t+\tau) \rangle},
\end{equation}

where $\mathbf{x}$ represents position, $\mathbf{r}$ is the radial separation vector with $r = |\mathbf{r}|$, and $\tau$ is the time lag (or lead). The angle brackets $\langle\cdot \rangle$ denote a time average, while the overbar indicates the spatial average.

The space-time cross-spectrum is defined as 

\begin{equation}\label{eqn:12}
	S_{wc}(K,\omega) = \int \int R_{wc}(r,\tau) e^{-i (K r-\omega \tau)} dr d \tau,
\end{equation}

where  $K=(k^2 +l^2)^{1/2}$ is the magnitude of the horizontal wavenumber and $\omega$ is the angular frequency. We examine the spatial and temporal dependence of the tracer flux by plotting the real part of the cospectrum (Fig.~\ref{f9}a-c)

\begin{equation}\label{eqn:13}
	\Re (S_{wc}(K,\omega)) = \Re{\{\hat{w}^{*}\hat{c}\}}(K,\omega), 
\end{equation}

with $\hat{w}$, and $\hat{c}$ denoting the Fourier transform of $w'$ and $c'$, and the operator $*$ representing the complex conjugate. When the total covariance $\int \int \Re (S_{wc}(K,\omega)) dK d\omega $ (Fig.~\ref{f9}d-f) is positive, it indicates that $w'$ and $c'$ are in phase at a particular frequency and wavenumber, while negative covariance indicates that $w'$ and $c'$ are out of phase at that spatio-temporal scale. 

To quantify the tracer flux as a function of wavenumber and frequency, we compute isotropic tracer covariance spectra at 40~$m$ depth, approximately where the horizontally averaged vertical tracer flux is maximum (the methodology is described in the SI, Section 5). Prior to the spectral analysis, area means were removed (excluding 48~$km$ from the boundaries in the meridional direction), temporal means were subtracted, and anomalies were tapered with a Tukey window ($r=0.1$) in space and time. 

At 40~$m$ depth, the front-only case shows predominantly negative covariance amounting to a downward tracer flux at subinertial frequencies and negligible contribution from inertial and super-inertial frequencies. When waves are included, the subinertial negative covariance (downward flux) increases, particularly for the F+SW case. NIWs induce both positive and negative covariances, with a net negative covariance (downward tracer flux) in the near-inertial frequency band. The subinertial covariance is largest at the 5-10~$km$ length scale, while the inertial (wave) contributions to the flux occur at larger scales (SI Fig.~S11). 

All three cases exhibit negative tracer covariance in the subinertial band, with additional positive contributions at higher frequencies at 40~$m$ depth. In the inertial band, the wave-forced cases show enhanced negative covariance across the mesoscale–submesoscale range, with patches of systematic positive contributions. At super-inertial frequencies, alternating negative and positive patterns emerge, though their integrated effect is small compared to subinertial and inertial frequencies. These results suggest that, below the mixed layer, vertical tracer transport arises from a combination of the background frontal flow and NIWs that interact with submesoscale currents.

\section{Conclusions}
\label{sec:4}

In this study, we investigated the vertical transport of passive tracers in an idealized ocean model, focusing on the influence of near-inertial wind forcing on a density front representative of Balearic Sea conditions during the CALYPSO campaign (February–March 2022). Using simulations at 1~$km$ horizontal resolution with varying NIWs, we find that tracer subduction is enhanced in the presence of NIWs, with the strongest wave-forced case producing a tracer deficit nearly twice that of the unforced front. PDFs reveal enhanced skewness in vorticity and negative skewness in divergence, consistent at submesoscales. The isopycnal slope exhibits pronounced extrema relative to the F case, indicating that frontogenesis becomes increasingly important -- particularly in the later stages of the simulation -- and is further enhanced in the strong wave-forced cases. 

Spectral analysis of vertical velocity reveals that inertial wind forcing modifies the power distribution across scales, particularly in the subinertial range, with multiple superinertial frequency bands forming harmonics of the inertial frequency. Nonlinear transfers emerge in the inertial and superinertial ranges, and tracer flux diagnostics show that subinertial components dominate vertical transport within the mixed layer, increasing with stronger forcing. Below the mixed layer, the tracer fluxes are modified at both the subinertial and inertial frequencies, and the net transport is greater than the transport contributed by fronts. In contrast, high-frequency motions at $2f, 3f, \ldots$ contribute little to net transport, as NIWs decay rapidly within the mixed layer, with only a fraction of energy penetrating below.

Our experiments employ idealized viscosity coefficients that do not account for enhanced mixing by near-inertial wave shear at the base of the mixed layer, which would likely increase vertical transport \citep{wang2025parameterization}. The results indicate that tracer transport by near-inertial pumping is not entirely reversible and that waves contribute to tracer transport through rectification in a submesoscale flow.

\bibliography{references.bib}

\section*{Open Research}

The PSOM model configuration can be found in the GitHub repository (\sloppy{\url{https://github.com/niharpaul/PSOM-NIWs-CALYPSO---JPO-2026}}). The datasets and analysis code associated with the Figures used are available at Zenodo (\mbox{\url{http://doi.org/10.5281/zenodo.14279472}}). Additionally, the three-dimensional data can be shared upon request to the corresponding author. The authors declare no conflicts of interest, and the results have not been previously published in any journal. 
	
	\section*{Acknowledgments}

This research was a part of the ``Coherent Lagrangian Pathways from the Surface Ocean to the Interior'' (CALYPSO) Department Research Initiative, supported by the US Office of Naval Research (ONR). NP acknowledges additional support from the National Aeronautics and Space Administration (NASA) Sub-Mesoscale Ocean Dynamics Experiment (SMODE) through Dr. J. Thomas Farrar. NP gratefully thanks Dr. J. Thomas Farrar, Dr. Eric D’Asaro, Dr. Michael Spall, Dr. Jai Sukhatme, and Dr. Dipanjan Chaudhuri for their valuable feedback and discussion on the analysis. The authors also acknowledge the attendees who provided comments on the poster presented at the 2024 Ocean Sciences Meeting, the 2024 Gordon Research Conference, and the 2025 Ocean Transport and Eddy Energy Climate Process Team (CPT) meeting at the Courant Institute of Mathematical Sciences, New York University, US.

\section*{Supplementary Information}

\begin{enumerate}
	\item SI text S1-S5.
	\item Tables S1.
    \item Movie S1.
   	\item Figures S1 to S11.
\end{enumerate}

\clearpage

\textbf{SI text S1:~Initialization of the density front}

Observations collected during the CALYPSO field campaign (Fig.~\ref{S1}) are used to set up the initial conditions in the model. The initialization of a front in the upper 200~$m$ of the channel domain is described by
\begin{align}
	&y_f(x)=\frac{L_y}{2}+Asin\left( \frac{2\pi x}{L_x}\right);\: A=1,\label{B1}\\
	& \Gamma(x,y)=0.5 \left[ 1+tanh \left( 2\pi B(y-y_f(x) \right) \right] ;\: B=0.4,\label{B2}\\
	&\sigma_\theta(x,y,z) =\sigma_{\theta s}(z) \left( 1-\Gamma \right) +\sigma_{\theta n}(z) \Gamma.\label{B3}
\end{align}
Here, $x,y$ are the zonal (along-channel) and meridional (across-channel) coordinates, and $y_f$ is the position of the center of the front. The channel dimensions $L_x,\: L_y$ are 96~$km$ and 192 $km$, respectively, representing the zonal and meridional extent of the domain. A sinusoidal meander of amplitude $A$ and wavelength $L_x$ is imposed on the frontal position (Equation \ref{B1}). The parameter $B$ controls the meridional width of the frontal zone where the potential density transitions from $\sigma_{\theta s}$ in the south (lighter side of the front) to $\sigma_{\theta n}$ in the north (denser side of the front). The initial $\sigma_\theta$ for a central value of $x$ is plotted in Fig.~\ref{S2}a, and the vertical profiles of $\sigma_{\theta s}$ and $\sigma_{\theta n}$ are shown in Fig.~\ref{S2}b,c. Below 200~$m$, there are no horizontal gradients, and the density stratification is based on an observed density profile from an ARGO float.\\ 

\textbf{SI text S2:~Generation of near-inertial waves}

In order to generate inertial oscillations, the front is forced with a time-varying windstress from $day$~5 to 8 (3 $days$) that rotates clockwise for 4 inertial periods of 0.75 $day$. The zonal and meridional windstress $\tau_x,\tau_y$ are shown in Fig.~\ref{S3}. The amplitude of the windstress is gradually ramped up and down in time by multiplying it by a Hanning window. The windstress is applied only to the central part of the channel domain to avoid wall effects.\\

\textbf{SI text S3:~Helmholtz Decomposition}

The Helmholtz decomposition separates the flow field into rotational and irrotational components.
Near the surface (3.5 $m$ depth in the model), the area-averaged rotational kinetic energy (KE) is elevated in the wave-forced cases compared to the F-only run immediately after inertial forcing, and accompanied by a strong divergent KE signal that gradually decays over the 60-$day$ integration (SI Fig. \ref{S4}). Notably, during the final 20~$days$, rotational and divergent KE remain nearly an order of magnitude higher in the wave-forced cases, suggesting that there is an exchange of energy between the balanced and unbalanced components of the circulation \citep{capet2008mesoscale,olbers2017closure,thomas2020turbulent}. External forcing at low wavenumber governs the exchange of energy between the balanced and unbalanced components of the flow through wave–mean flow interactions and nonlinear turbulent processes \citep{gage1985spectrum,maltrud1991energy,danilov2004scaling}. The resulting envelope is strongly scale-dependent, delineating the transition between slow and fast dynamics as the simulation evolves (SI Fig. \ref{S4}). The H\"{o}vm\"{o}ller diagrams of horizontal and vertical KE in SI Fig. \ref{S5} and \ref{S6} indicate that wave forcing enhances energy in the upper 100 $m$.\\

\textbf{SI text S4:~Kinetic energy-wavenumber isotropic (radial) spectra}

We adopt the methodology described in \cite{errico1985spectra,durran2017practical,samelson2024models} and briefly illustrate it here. Let $k_x$, $k_y$, and $\hat{u}(k_x,k_y)$, $\hat{v}(k_x,k_y)$ be the dimensional wavenumbers in the $x$ and $y$ directions, and the two-dimensional Fourier transforms of the velocity field. The horizontal wavenumber is defined by $K=(k_x^2+ k^2_y)^{1/2}$ with the integral of the spectral density of the two-dimensional kinetic energy (KE) $E(K)$ equal to the integral of the KE averaged over the domain given by Equation \ref{C1},\\

\begin{equation}\label{C1}
	\left(\frac{\overline{\mathbf{u}\cdot \mathbf{u}}}{2}\right)=\int_0^{\infty}E(K)dK,
\end{equation}

Let $\mathbf{c}(\theta)=(Kcos\theta,Ksin\theta)$ be represented in its parametric form; therefore, the $E(K)$ along the radial direction can be written as

\begin{equation}\label{C2}
	E(K)=\frac{1}{2}\int_0^{2\pi} \left\{\hat{u}[\mathbf{c}(\theta)]\hat{u}^{*}[\mathbf{c}(\theta)] \right\}+\left\{\hat{v}[\mathbf{c}(\theta)]\hat{v}^{*}[\mathbf{c}(\theta)] \right\}  K d\theta.
\end{equation}

The $E(K)$ from the horizontal velocities ($u_{r,s}$, $v_{r,s}$) is computed on the periodic mesh over the points of the grid $N_x$, $N_y$ along x and y, given by

\begin{align}\label{C3}
	&x_r=(r-1)\Delta x,\: r=1,2,...,N_x,\\
	&y_s=(s-1)\Delta y,\: s=1,2,...,N_y.
\end{align}

Then we use the Parseval theorem to convert from the discretized average KE to its two-dimensional discrete Fourier transform form in Equation \ref{C4},

\begin{equation}\label{C4}
	\frac{1}{N_xN_y}\sum_{r=1}^{N_x}\sum_{s=1}^{N_y} u^2_{r,s}+v^2_{r,s}=\sum_{l=1}^{N_x} \sum_{m=1}^{N_y}{\hat{u}^{}}_{l,m}{\hat{u}^{*}}_{l,m}+\hat{v}^{}_{l,m}{\hat{v}^{*}}_{l,m}, 
\end{equation}

Using the length of the domain as $L_x=N_x \Delta x$ and $L_y= N_y \Delta y$, we obtain $\Delta k_x=2\pi/L_x$ and $\Delta k_y=2\pi/L_y$; therefore, Equation \ref{C4} becomes

\begin{equation}\label{C5}
	\frac{1}{L_xL_y}\sum_{r=1}^{N_x}\sum_{s=1}^{N_y}\frac{u^2_{r,s}+v^2_{r,s}}{2}\Delta x\Delta y=\frac{L_xL_y}{8\pi^2}\sum_{l=1}^{N_x}\sum_{m=1}^{N_y}(\hat{u}^{}_{l,m}\hat{u}^{*}_{l,m}+\hat{v}^{}_{l,m}\hat{v}^{*}_{l,m})\Delta k_x \Delta k_y.
\end{equation}

We discretize the $2D$ wavenumber in multiples of
the maximum one-dimensional wavenumber, $\Delta K=max(\Delta k_x, \Delta k_y)$, such that

\begin{equation}\label{C6}
	k_p=p\Delta K,\: p=1,2, ..., N_{max},
\end{equation}

where, $N_{max}=[\sqrt{2}max(N_x/2, N_y/2)]$. We then define $R(p)$ as the set of wavenumber indices ($l$, $m$) that satisfy the equation \ref{C8}, and plot the spectrum $\tilde{E}(k_p)$ through the wavenumber $k_p=N\Delta K/2$.

\begin{align}
	&k_p-\Delta K/2\le (k^2_{x_l}+k^2_{y_m})^{1/2}<k_p+\Delta K/2,\label{C7}\\
	&\tilde{E}(k_p)=\frac{L_xL_y min(\Delta k_x, \Delta k_y)}{8\pi^2}\sum_{l,m \in R(p)}(\hat{u}^{}_{l,m}\hat{u}^{*}_{l,m}+\hat{v}^{}_{l,m}\hat{v}^{*}_{l,m}).\label{C8}
\end{align}

\textbf{SI text S5:~Frequency-wavenumber isotropic (radial) spectra}

The frequency-horizontal wavenumber isotropic spectral density has been calculated from three three-dimensional discrete Fourier transforms following the methodology adopted from \cite{savage2017spectral}. The three-dimensional Fourier transform of a variable $\eta$ along the dimension corresponding to the indices $q$, $g$, and $n$ given by

\begin{equation}\label{D1}
	\hat{\eta}_{p,h,m}(k_p,l_h,\omega_m)=\sum_{q=0}^{Q-1}\sum_{g=0}^{G-1}\sum_{n=0}^{N-1}\eta_{q,g,n} e^{-2\pi i({\frac{pq}{Q}+\frac{hg}{G}+\frac{mn}{M}})},
\end{equation}

where, $\hat{\eta}$ is the Fourier transform corresponding to the indices $p$, $h$, and $m$ as a function of $k_p$, $l_h$, and $\omega_m$ denoting the zonal, meridional wavenumber, and frequency. The length of the physical samples in the zonal, meridional, and temporal directions is given by $Q$, $G$, and $N$, respectively. 

The spectral density $|\hat{\eta}(K,\omega)|^2$ associated with the $r$-th element of the isotropic wavenumber, $k_r$ is given, 

\begin{equation}\label{D2}
	|\hat{\eta}_r(K_r,\omega)|^2=
	\begin{cases}
		\frac{2}{QGN}\left[ \sum_{p=1}^{\chi}\sum_{h=1}^{\zeta}|\hat{\eta}_{p,h}(k_p,l_h,\omega)|^2\right],& \text{if}\: r= 1,\\
		\frac{2}{QGN}\left[\sum_{p=1}^{\chi}\sum_{h=1}^{\zeta}|\hat{\eta}_{p,h}(k_p,l_h,\omega)|^2\right]-\sum_{\gamma=1}^{r-1}|\hat{\eta}_{\gamma}(k_\gamma,\omega)|, & \text{if}\: r>1.
	\end{cases}
\end{equation}

where $k^2_{\chi}+l^2_{\zeta}$, and $r$ is an index that spans 1 to the length of $K$ and $|\hat{\eta}(K,\omega)|^2$ is computed iteratively. The first term on the right-hand side of the equation is a sum over all values of $|\hat{\eta}(k,l,\omega)|$ for which $p$ and $h$ satisfy condition $k^2_\chi+l^2_\zeta<K^2_r$, and the second term on the right-hand side in the condition that $r>1$ is a sum over all previously computed values of $\hat{\eta_\gamma}(K_r,\omega)$. The method conserves variance while transforming from anisotropic to isotropic spectral density.\\
\clearpage
\textbf{Supplementary Table S1}

\setcounter{table}{0}
\renewcommand{\thetable}{S\arabic{table}}
\begin{table}[!h]
	\begin{center}
		\begin{tabular}{ |p{4cm}||p{3cm}|p{3cm}|p{3cm}|}
			\hline
			Barotropic/baroclinic wave speed ($m$ ${s^{-1}}$)& F & F+WW & F+SW\\
			\hline
			$c_0$ & 98.2124 & 98.2124  & 98.2124\\
			\hline
			$c_1$ & 0.4822  & 0.4811 & 0.4801\\
			\hline
			$c_2$ & 0.2635  & 0.2608 & 0.2581\\
			\hline
			$c_3$ & 0.1656  & 0.1651 & 0.1645\\
			\hline
			$c_4$ & 0.1298  & 0.1283 & 0.1274\\
			\hline
			$c_5$ & 0.1027  & 0.1019 & 0.1013\\
			\hline
			$c_6$ & 0.0863  & 0.0852 & 0.0846\\
			\hline
			$c_7$ & 0.0733  & 0.0727 & 0.0722\\
			\hline
			$c_8$ & 0.0641  & 0.0634 & 0.0630\\
			\hline
			$c_9$ & 0.0570  & 0.0564 & 0.0560\\
			\hline
			$c_{10}$ & 0.0512 & 0.0506 & 0.0503\\
			\hline
		\end{tabular}
		\caption{Barotropic ($c_0$) and baroclinic wave speeds ($c_m$, $m=1,2,...,10$) for the three experiments: F, F+WW, and F+SW.}
	\end{center}
\end{table}

\textbf{Supplementary Movies}

\textbf{Movie S1}: : (a) Free evolution of a density front (F). (b) Front with strong-wave (F+SW) forcing (inertial stress $|\tau|_{max}=0.5$ $Pa$). (c) No-front case with the initial stratification following the lighter side of the front and subjected to the same strong-wave (SW) forcing (
$|\tau|_{max}=0.5$ $Pa$), with potential-density contours overlaid at a depth of 3.5 $m$. For the forced simulations, windstress is applied from days 5–8 and ramped up and down using a Hanning window. The tracer is initialized on $day$~8 when the wind forcing is turned off. The quantity $D$ in \% denotes the percentage of tracer deficit relative to its initial distribution on $day$~8 and is computed and shown for the upper 200~$m$.\\

\setcounter{figure}{0}
\renewcommand{\figurename}{Fig.}
\renewcommand{\thefigure}{S\arabic{figure}}
\begin{figure}
	\centering 
	\includegraphics[width=\textwidth]{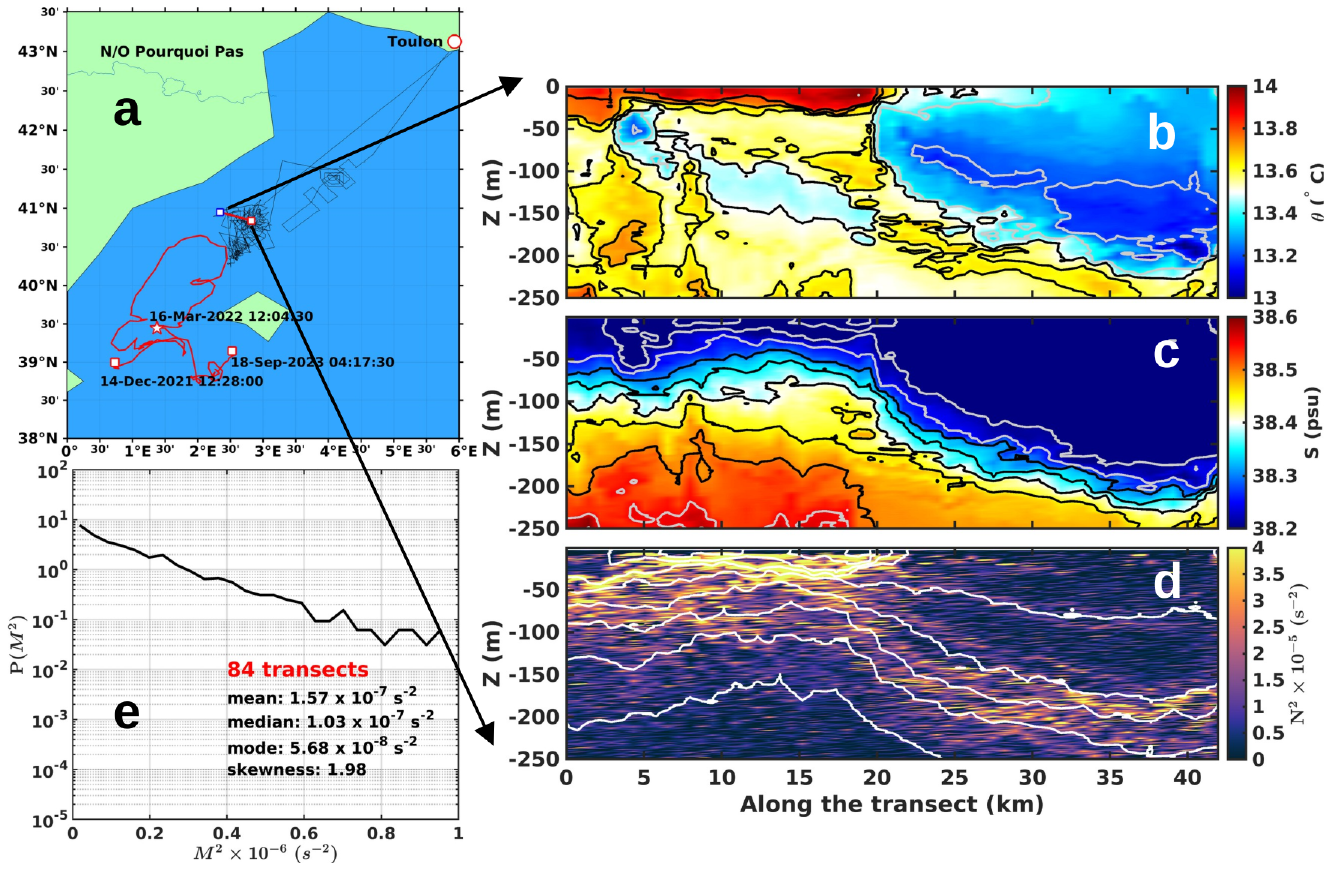}
	\caption{(a) Survey of the vessel {\it N/O Pourquoi Pas} conducted in the Balearic Sea from February to March 2022. The trajectory of the ARGO float is also shown, spanning December 14, 2021, to September 18, 2023. (b-d) Potential temperature ($\theta$, $^\circ C$), salinity ($S$, $psu$), and squared Brunt-V\"{a}is\"{a}l\"{a} frequency ($N^2$, $s^{-2}$) with contours of potential density anomaly ($\sigma_\theta$, $kg$ $m^{-3}$) across the mesoscale front along the transect highlighted in red in (a). (e) Probability density function of $M^2$, defined as $|\nabla_H b|$, at the surface from 84 transects, where $b$ is buoyancy.}
	\label{S1}
\end{figure}

\begin{figure}
	\includegraphics[width=\textwidth]{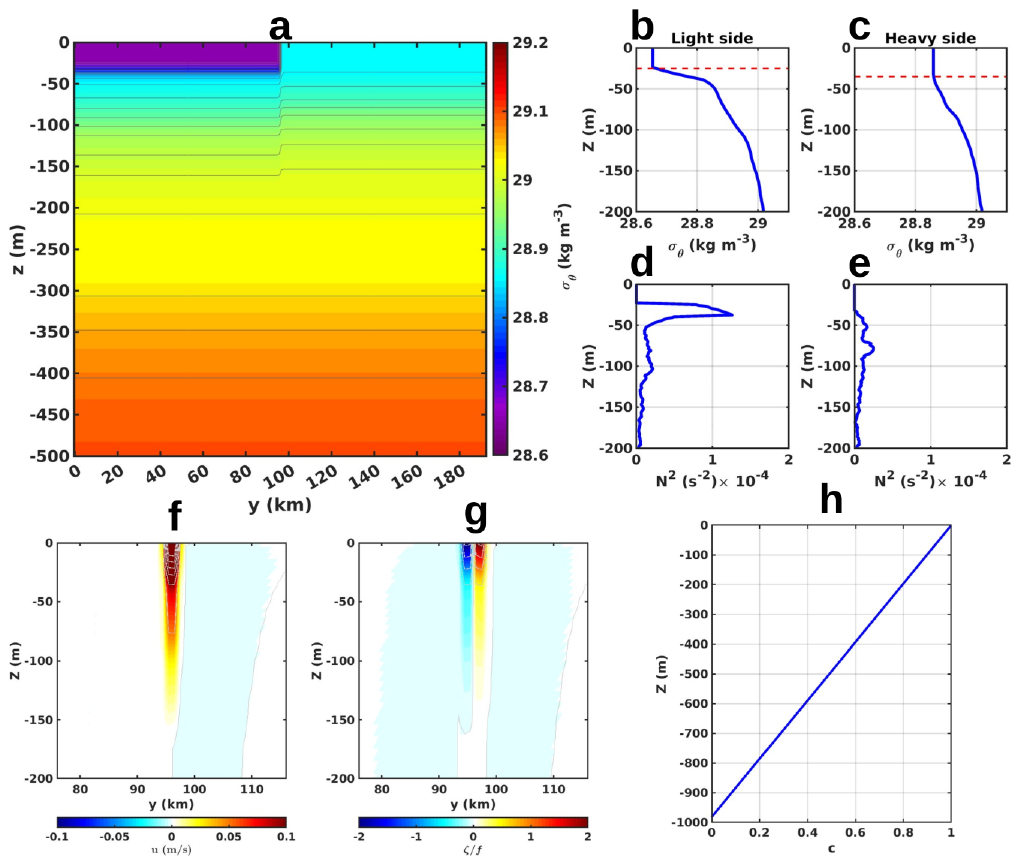}
	\caption{(a) Potential density anomaly ($\sigma_{\theta}$) of the front associated with the transect shown in Fig. \ref{S1} (upper 500 $m$). (b--e) $\sigma_\theta$ and squared Brunt--V\"{a}is\"{a}l\"{a} frequency ($N^2$) with tightness = 0.4, $M^2 = 9.023\times10^{-7}$ $s^{-2}$, and $M^2/f^2 = 101.13$, where $f$ is the Coriolis parameter. Red dashed lines indicate the mixed layer depths on the light (25 $m$) and heavy (35 $m$) sides of the front. (f--h) Initial conditions: zonal velocity $u$ ($m$ $s^{-1}$), Rossby number $\zeta/f$ in the upper 200 $m$, and a linear tracer profile $c$ ($mg$ $m^{-3}$) versus depth, varying from 0 at the bottom to 1 at the surface.}
	\label{S2}
\end{figure}

\begin{figure}
	\includegraphics[width=\textwidth]{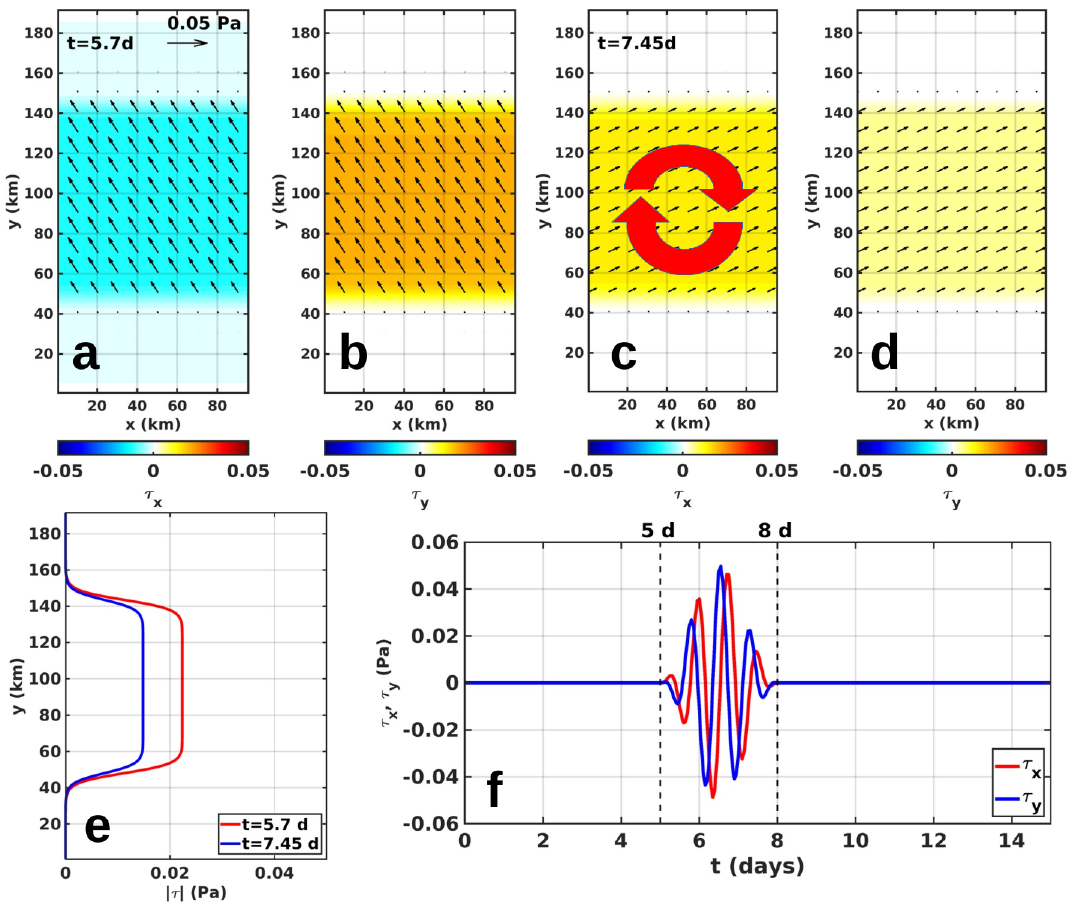}
	\caption{(a,b) and (c,d) show spatial maps of zonal and meridional windstress with quivers ($\tau_x$,~$\tau_y$) at $t = 5.7~day$ and $t = 7.45~day$, respectively. 
		(e) shows the meridional profile of windstress magnitude corresponding to the times shown in (a--d). (f) presents the 15-$day$ time series of surface windstress ($\tau_x$,~$\tau_y$) at the domain center. Winds are switched on and off between days 5--8 to generate near-inertial waves (NIWs). Clockwise arrows in (c) indicate the rotation direction of the inertial winds.}
	\label{S3}
\end{figure}

\begin{figure}
	\includegraphics[width=\textwidth]{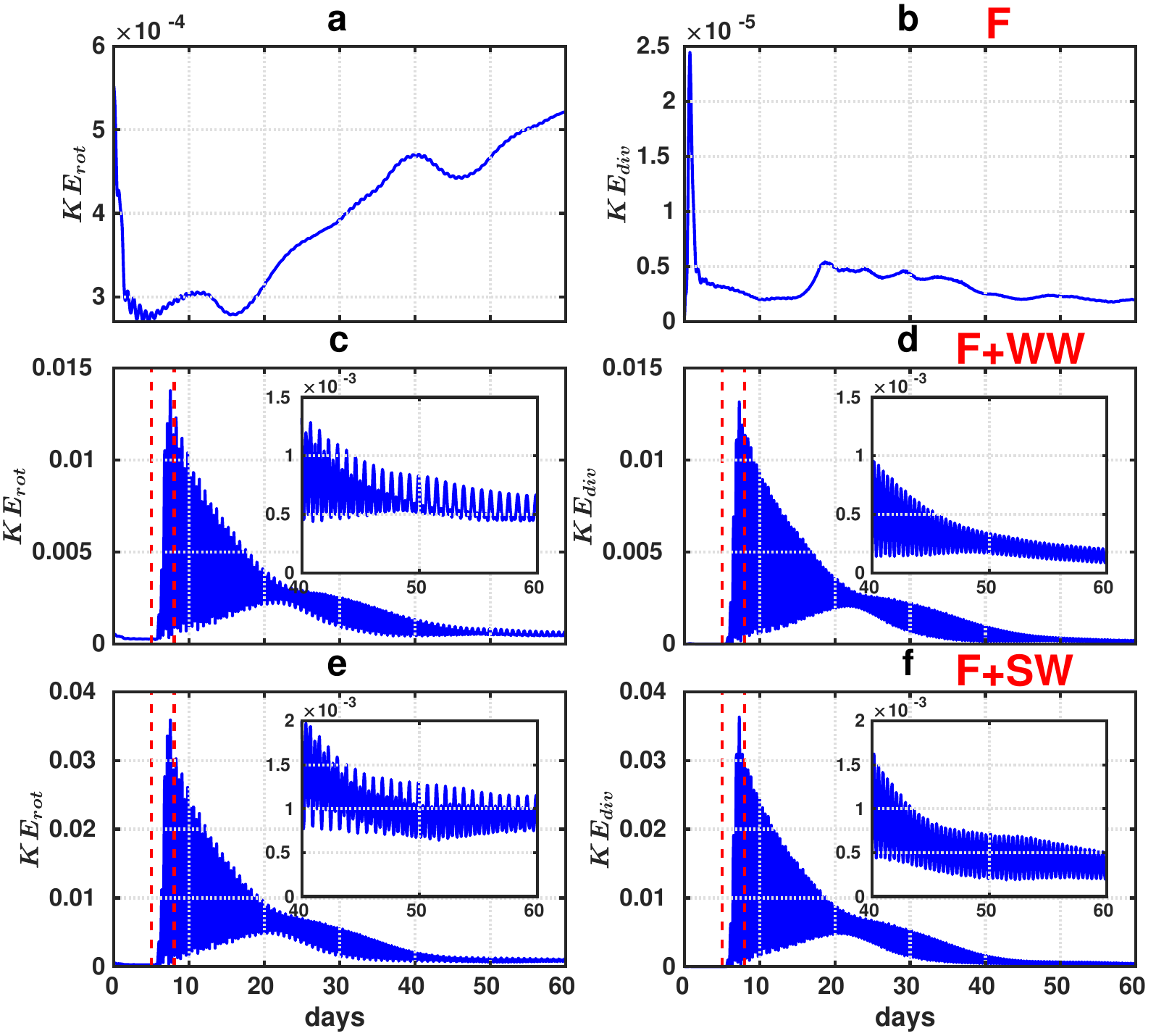}
	\caption{The left column: panels (a), (c), and (e), display the rotational kinetic energy ($KE_{rot}$), while the right column: panels (b), (d), and (f), represent the divergent kinetic energy ($KE_{div}$) for the Front (F) and the front with weak and strong waves (F+WW, F+SW) at a depth of 3.5 $m$ throughout the 60-$day$ simulation. Insets in panels (c)--(f) highlight the corresponding kinetic energy between $days$~40--60 of the simulation. Vertical dashed lines in panels 2 and 3 indicate $days$~5 and 8, respectively.}
	\label{S4}
\end{figure}

\begin{figure}
	\includegraphics[width=\textwidth]{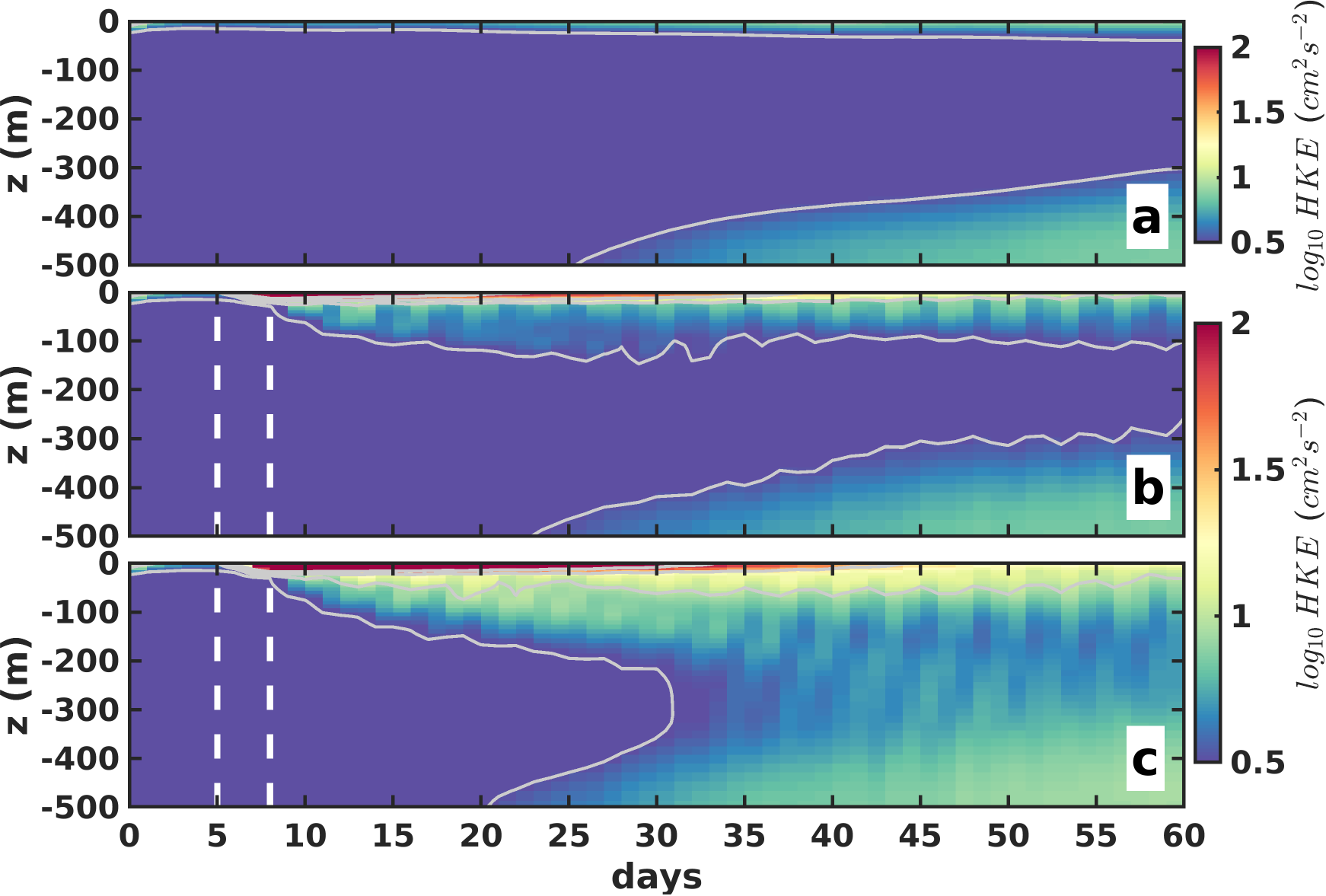}
	\caption{(a), (b), (c) shows the H\"{o}vm\"{o}ller of average of horizontal kinetic energy (HKE) over the domain at per with the region of isospectra of the front (F), front with weak wave (F+WW), and strong wave cases (F+SW) for the upper 500 $m$. The white vertical dashed line shows the turning on ($day$~5) and off period ($day$~8) of the inertial wind.}
	\label{S5}
\end{figure}

\begin{figure}
	\includegraphics[width=\textwidth]{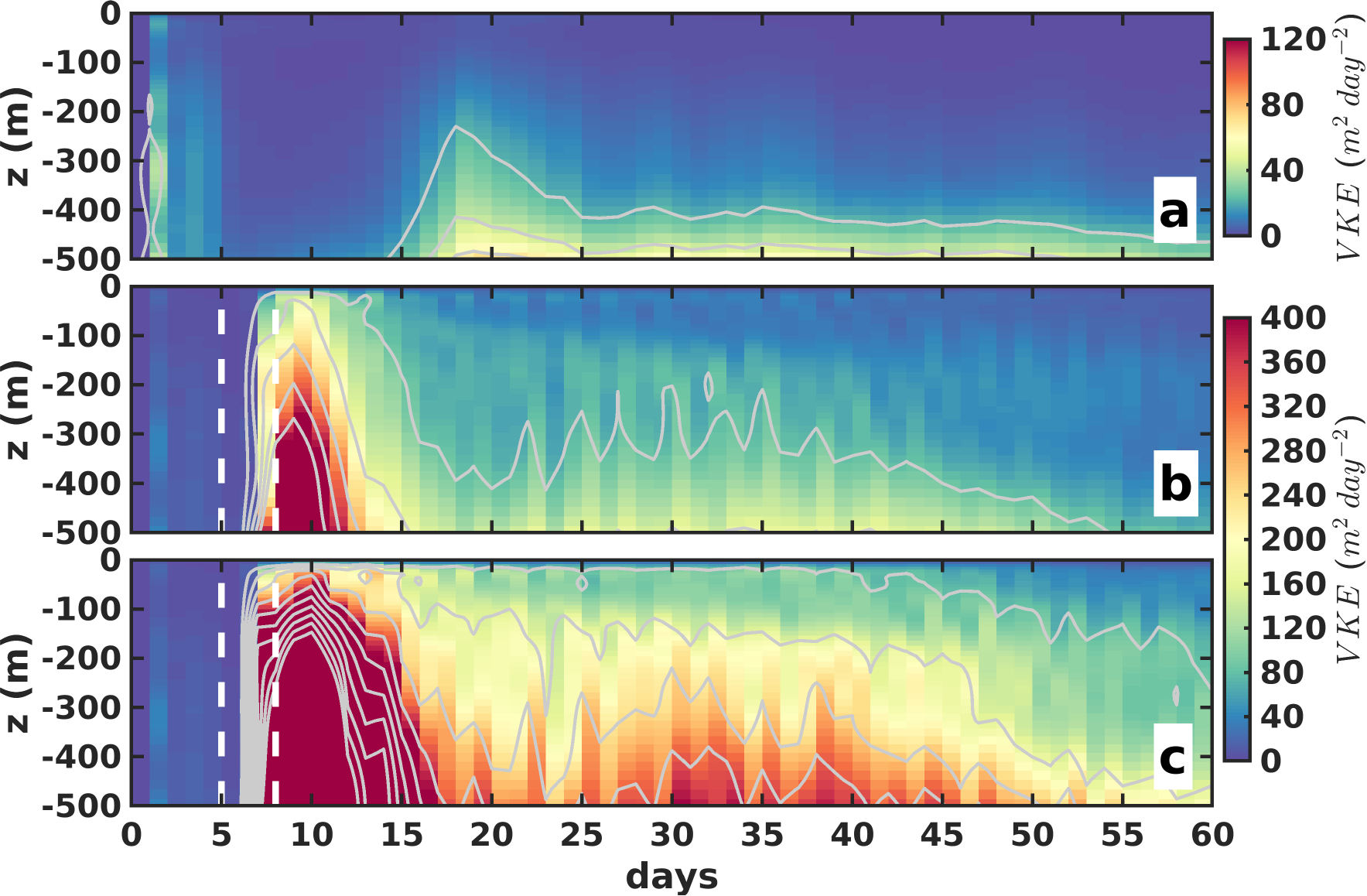}
	\caption{(a), (b), (c) shows the H\"{o}vm\"{o}ller of average of vertical kinetic energy (VKE) over the domain at per with the region of isospectra of the front (F), front with weak wave (F+WW), and strong wave cases (F+SW) for the upper 500~$m$. The white vertical dashed line shows the turning on ($day$~5) and off period ($day$~8) of the inertial wind.}
	\label{S6}
\end{figure}

\begin{figure}
	\includegraphics[width=\textwidth]{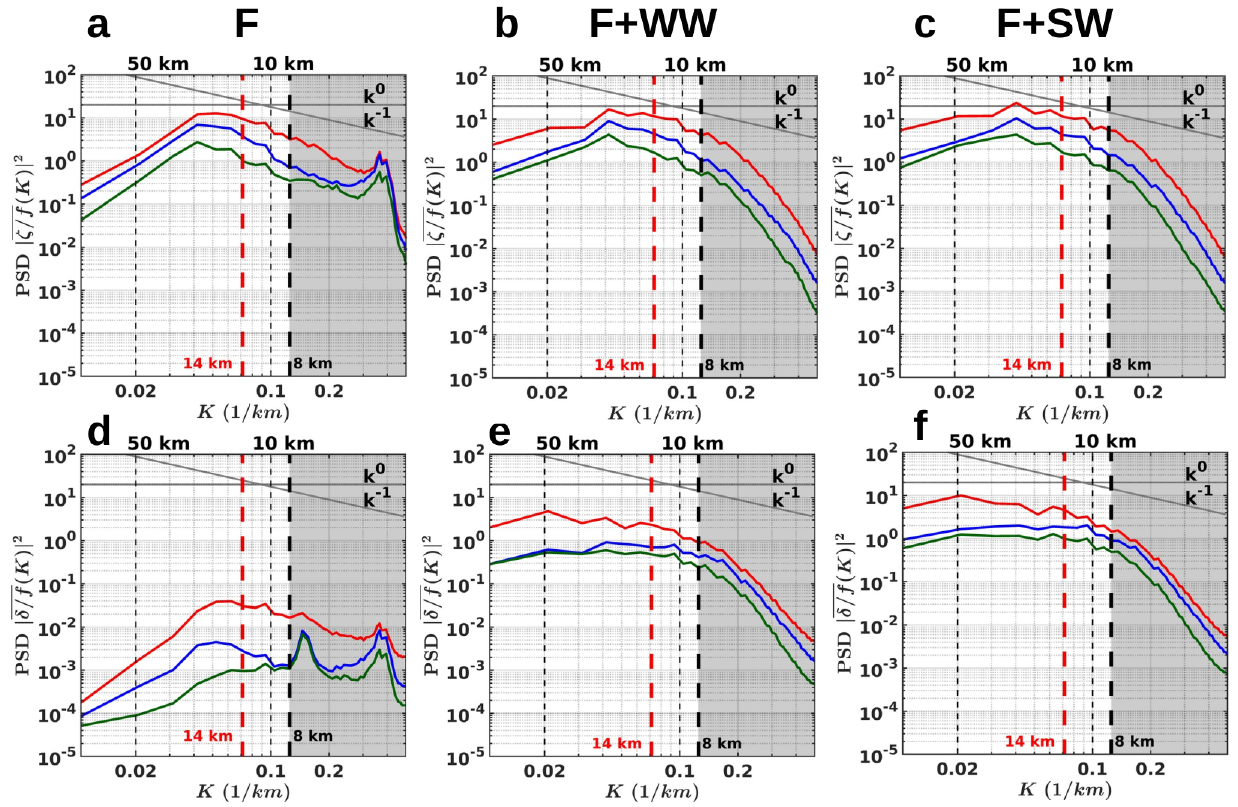}
	\caption{(a--c) and (d--f) show the isotropic power spectral density of the normalized vorticity($\zeta/f$) and divergence ($\delta/f$), respectively, as a function of horizontal wavenumber 
		$\boldsymbol{K} = \sqrt{k_x^2 + k_y^2}$, where $k_x$ and $k_y$ are the zonal and meridional wavenumbers 
		($\boldsymbol{K}$ in units of $km^{-1}$). Results are shown for the three cases--Front (F), Front with weak waves (F+WW), and Front with strong waves (F+SW)--over the period from $day$~8 to $day$~60 at depths of 3.5, 20, and 60 $m$. The gray lines in (a--f) indicate spectral slopes of~0 and~$-1$. The shaded region marks the unreliable part of the spectra ($\boldsymbol{K} \le 8$~$km$). The Rossby radius of deformation, $L_R = 14$~$km$, is indicated in red, as discussed in the text.}
	\label{S7}
\end{figure}

\begin{figure}
	\includegraphics[width=\textwidth]{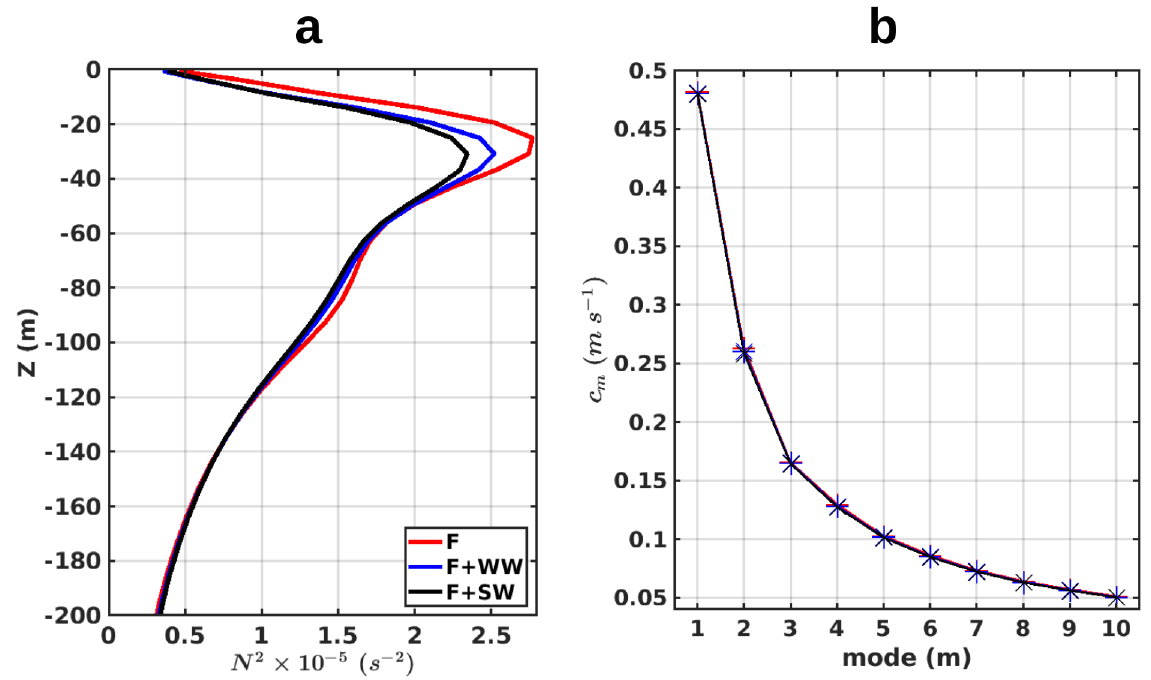}
	\caption{(a) Time-mean squared buoyancy frequency ($N^2$) averaged over $days$~8--60. (b) Baroclinic phase speeds ($c_m$) for the first ten modes, compared among the F, F+WW, and F+SW experiments.}
	\label{S8}
\end{figure}

\begin{figure}
	\includegraphics[width=\textwidth]{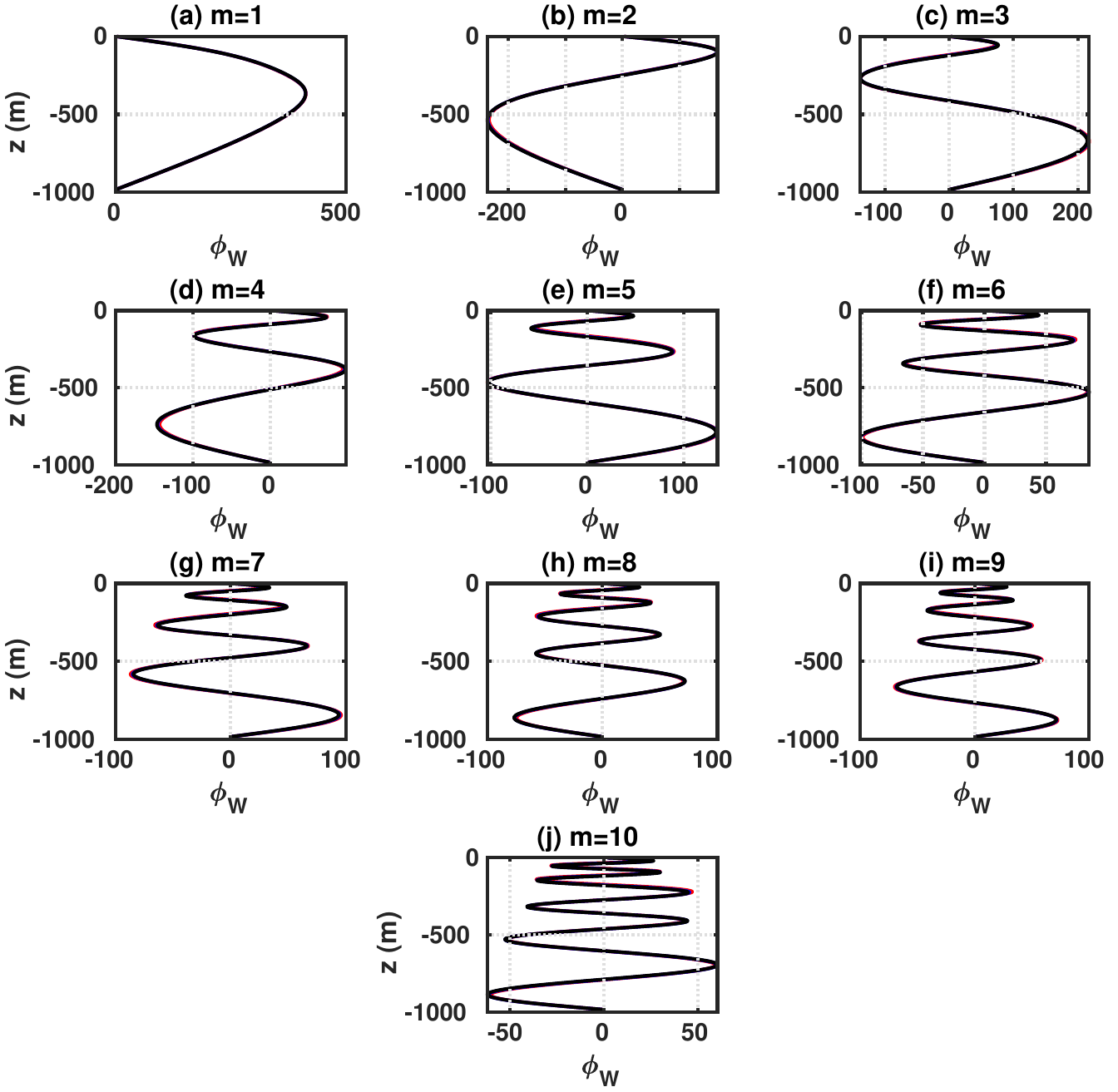}
	\caption{Panels (a–j) display the eigenfunctions of the vertical velocity, $\phi_w$, as a function of depth for modes $m=1$ to $m=10$. Results are shown for the three experiments: F (red), F+WW (blue), and F+SW (black).}
	\label{S9}
\end{figure}

\begin{figure}
	\includegraphics[width=\textwidth]{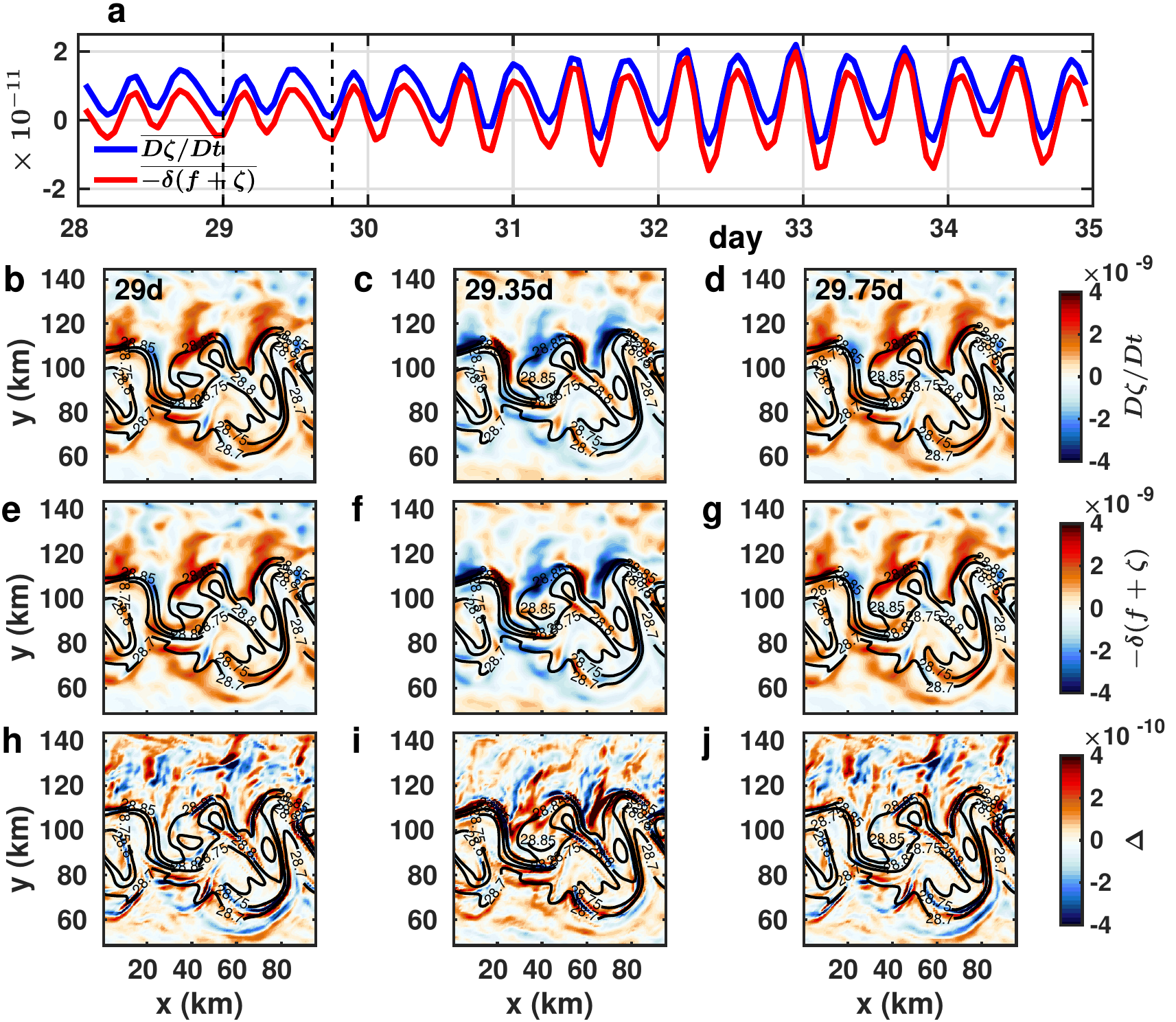}
	\caption{(a) Domain-averaged $\overline{D\zeta/Dt}$ and $-\overline{\delta(f+\zeta)}$ as functions of time from $day$~28 to $day$~35 at a depth of 3.5~$m$ for the SW simulation. (b–d), (e–g), and (h–j) show spatial maps of $D\zeta/Dt$, $-\delta(f+\zeta)$, and their difference $\Delta$ (panel 2 minus panel 3), respectively, over one inertial period at $days$~29, 29.35, and 29.75, with potential density contours overlaid. Dashed vertical black lines in panel (a) indicate the inertial period corresponding to the spatial maps.}
	\label{S10}
\end{figure}

\begin{figure}
	\includegraphics[width=\textwidth]{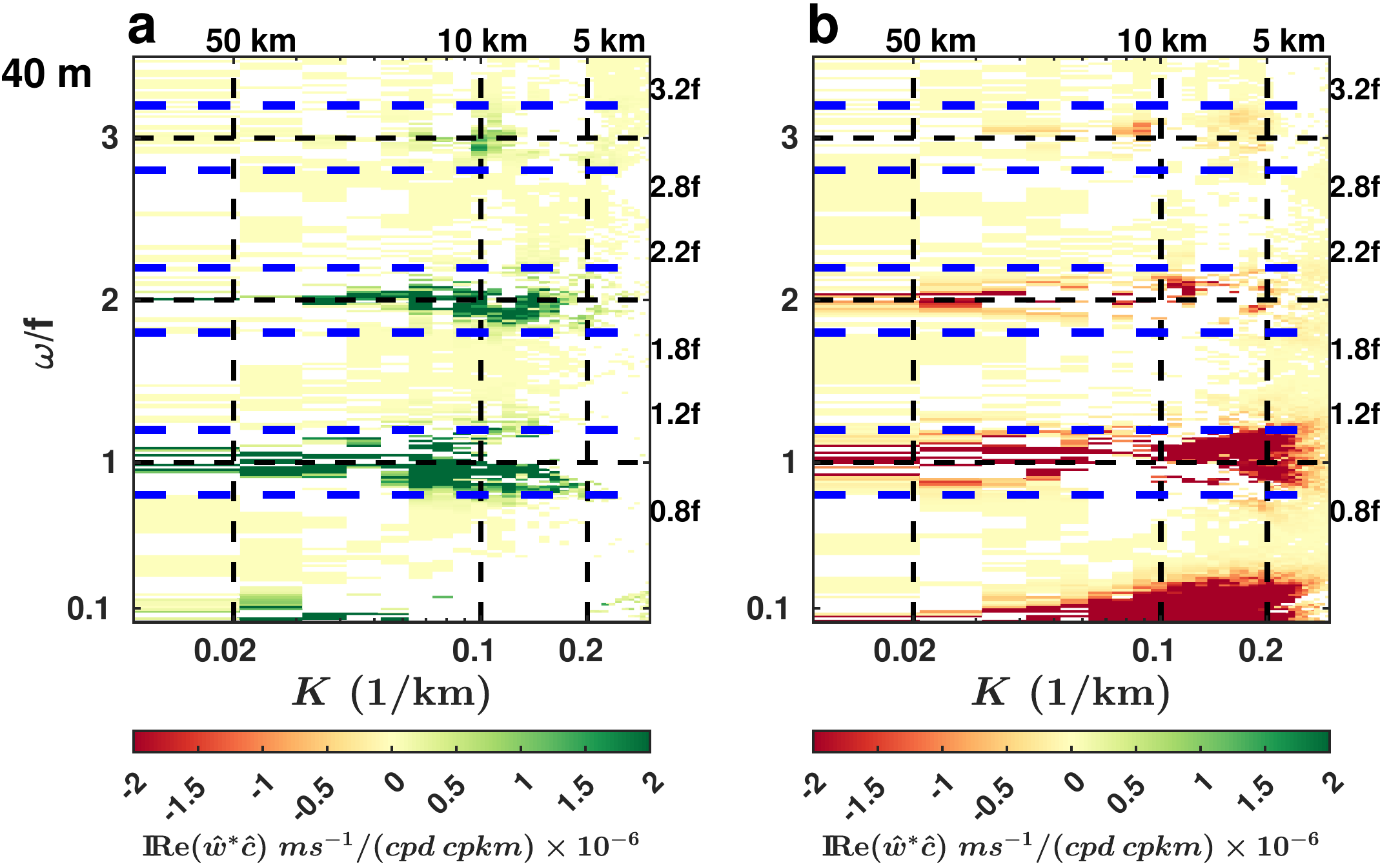}
	\caption{(a) shows the positive components, and (b) the negative components, of the frequency–wavenumber isospectra of tracer covariances at a depth of 40~$m$ for the F+SW experiment. Blue dashed lines denote the inertial and superinertial frequency bands ($f$, $2f$, and $3f$), each defined by an interval of $\pm,0.2f$ about the central frequency.}
	\label{S11}
\end{figure}

\end{document}